\documentclass[journal,twocolumn,10pt,twoside]{IEEEtranTCOM}
\normalsize

\ifCLASSINFOpdf
   \usepackage[pdftex]{graphicx}
   \graphicspath{{../pdf/}{../jpeg/}}
   \DeclareGraphicsExtensions{.pdf,.jpeg,.png}
\else
\fi

\usepackage[cmex10]{amsmath}

\usepackage{multicol}
\usepackage{float}
\usepackage{amsmath}
\usepackage{amssymb}
\usepackage{paracol}
\usepackage{multirow}
\usepackage[nolist,nohyperlinks]{acronym}
\usepackage{dblfloatfix} 
\usepackage[table]{xcolor}
\usepackage{array} 

\usepackage{eso-pic}

\newcolumntype{L}[1]{>{\raggedright\arraybackslash}m{#1}}
\newcolumntype{C}[1]{>{\centering\arraybackslash}m{#1}}
\newcolumntype{R}[1]{>{\raggedleft\arraybackslash}m{#1}}

\makeatletter
\newcommand{\acx}{\protect\@acx}%
\newcommand{\@acx}[1]{%
  \ifAC@dua
   \acl{#1}%
  \else
   \expandafter\ifx\csname ac@#1\endcsname\AC@used
      \acs{#1}%
   \else
      \acl{#1}%
   \fi
  \fi
}
\makeatother

\begin{acronym}
\acro{SG}{Smart Grid}
\acro{AMR}{automatic meter reading}
\acro{AMI}{advanced metering infrastructure}
\acro{BW}{bandwidth}
\acro{PLC}{power line communications}
\acro{NB-PLC}{narrowband \acx{PLC}}
\acro{OFDM}{orthogonal frequency division multiplexing}
\acro{DFT}{discrete Fourier transform}
\acro{IDFT}{inverse \acx{DFT}}
\acro{ICI}{intercarrier interference}
\acro{ISI}{intersymbol interference}

\acro{AIC}{active interference cancellation}
\acro{CC}{cancellation carriers}
\acro{OOBE}{out-of-band emission}
\acro{PSD}{power spectral density}
\acro{RC}{raised cosine}
\acro{RHS}{right hand side}
\acro{PAPR}{peak-to-average power ratio}
\acro{w.r.t.}{with respect to}

\acro{LTE}{long-term evolution}
\acro{4G}{fourth generation}
\acro{5G}{fifth generation}
\acro{IoT}{Internet of Things}
\acro{NB-IoT}{narrowband \acx{IoT}}
\acro{CR}{cognitive radio}
\acro{DSL}{digital subscriber line}
\acro{AST}{adaptive symbol transition}
\acro{FIR}{finite impulse response}
\acro{IIR}{infinite impulse response}
\acro{BER}{bit error rate}

\acro{ACLR}{adjacent channel leakage ratio}
\acro{CP-OFDM}{cyclic prefix based \ac{OFDM}}
\acro{FBMC}{filter bank multicarrier}
\acro{UFMC}{universal-filtered multicarrier}
\acro{BF-OFDM}{block-filtered \ac{OFDM}}

\end{acronym}

\DeclareMathOperator*{\argmin}{arg\,min} 

\definecolor{reviewer1}{rgb}{0 0 0}
\definecolor{reviewer2}{rgb}{0, 0, 0}

\begin{document}
\AddToShipoutPicture{%
  \put(0,\LenToUnit{\paperheight-1.8cm}){%
    \makebox[\paperwidth][c]{%
      \footnotesize
      \parbox{\textwidth}{
\vspace{-2.2cm}
© 2023 IEEE.  Personal use of this material is permitted.  Permission from IEEE must be obtained for all other uses,
 in any current or future media, including reprinting/republishing this material for advertising or promotional purposes,
 creating new collective works, for resale or redistribution to servers or lists, or reuse of any copyrighted component of this work in other works.
}

    }%
  }%
}
%
\title{Low-complexity spectral shaping method for OFDM signals with dynamically adaptive emission mask}

%
%
%

\author{Javier Giménez, José A. Cortés and Luis Díez
\thanks{The authors are with Communications and Signal Processing Lab, Telecommunication Research Institute (TELMA), Universidad de Málaga, E.T.S. Ingeniería de Telecomunicación, Bulevar Louis Pasteur 35, 29010 Málaga, Spain.
Corresponding author: Javier Giménez (e-mail: javierg@ic.uma.es). This work has been funded in part by the Spanish Government under project PID2019-109842RB-I00 and FPU grant FPU20/03782, by the European Fund for Regional Development (FEDER), Junta de Andalucía and the Universidad de Málaga under projects P18-TP-3587 and UMA20-FEDERJA-002.}}

\maketitle
\vspace{-1.75cm}

\begin{abstract}
\Ac{OFDM} signals with rectangular pulses exhibit low spectral confinement. Shaping their power spectral density (PSD) is imperative in the increasingly overcrowded spectrum to benefit from the cognitive radio (CR) paradigm. However, since the available spectrum is non-contiguous and its occupancy changes with time, the spectral shaping solution has to be dynamically adapted. This work proposes a framework that allows using a reduced set of preoptimized pulses to shape the spectrum of \ac{OFDM} signals, irrespective of its spectral width and location, by means of simple transformations. The employed pulses combine active interference cancellation (AIC) and adaptive symbol transition (AST) terms in a transparent way to the receiver. They can be easily adapted online by the communication device to changes in the location or width of the transmission band, which contrasts with existing methods of the same type that require solving NP-hard optimization problems.
\end{abstract}


%
\IEEEpeerreviewmaketitle

\section{Introduction}

\IEEEPARstart{O}{rthogonal} frequency division multiplexing (OFDM) is widely used in current wireless and wired communication systems. It features important advantages such as the ability to perform spectrum aggregation, robustness to frequency selectivity and low transceiver complexity, among others \cite{Zaidi2016}. However, rectangularly windowed \ac{OFDM} has low spectral confinement, which causes inefficient utilization of the spectrum and hinders a flexible distribution of the spectral resources, e.g., it obliges to devote $10$\% of the \ac{LTE} bandwidth to guardbands and causes inter-numerology interference in \ac{5G} systems \cite{Guan2017}\cite{Zhang2018}. 

The large \ac{OOBE} of \ac{OFDM} signals is also a drawback for the dynamic sharing of the spectrum, which is a key element to exploit unused spectral resources.  This is generally referred to as \ac{CR} \cite{Haykin05}, although it is also employed in wired systems such as \ac{PLC} to prevent its radiated emissions from interfering other wired services, such as \ac{DSL}, and wireless ones, like aeronautical mobile services, that might coexist in the same area \cite{Galli2016bis}\cite{G.9977}\cite{EN50561-1}. 


To overcome this issue, alternative multicarrier modulations like \ac{FBMC}, \ac{UFMC} \cite{Vakilian2013} and \ac{BF-OFDM} \cite{Gerzaguet2017} have been proposed. However, a large number of techniques have also been developed to reduce the \ac{OOBE} of \ac{OFDM} signals. Nulling carriers located at the edges of the transmission band is the simplest one, but it considerably penalizes the data rate. Hence, a myriad of time and frequency-domain methods with reduced impact in the system throughput have been proposed \cite{You14}. Filtering, pulse-shaping and \ac{AST} are typical time-domain methods which might be applied transparently to the receiver. Filtering with relatively low-order \ac{FIR} filters has been proposed for \ac{5G} \cite{Abdoli15}\cite{Pitaval2019}, but it is impractical for \ac{CR} applications because the available spectrum is usually non-contiguous and changes dynamically. 

Pulse-shaping smooths the symbol boundaries using a window with tapered transitions (e.g., raised cosine), while symbol transitions are adaptively smoothed in \ac{AST}. Both techniques can be applied in a receiver agnostic way at the cost of reducing the effective length of the cyclic prefix. While former versions of \ac{AST} oblige to solve a minimization problem for each symbol \cite{Mahmoud08}, proposals in which the optimization is performed offline have been recently made \cite{Hussain2020}. 

Among the frequency-domain methods, precoding has attracted great interest because of its considerable \ac{OOBE} reduction, which can be done without penalizing the spectral efficiency \cite{Chung08}\cite{vandeBeek09bis}\cite{Kumar2021}. However, the receiver must be aware of the precoding applied by the transmitter to avoid degrading the \ac{BER}. \Ac{AIC} is a frequency-domain strategy in which a set of carriers, referred to as \ac{CC}, are devoted to lower the \ac{OOBE}. \ac{AIC} can be applied transparently to the receiver at the cost of a minimum power and spectral efficiency penalty. The modulating values of the \ac{CC} are normally linearly derived from the ones that modulate the data carriers \cite{Brandes06}. The proposals in \cite{Schmidt13}\cite{Hussain2021} calculate the combination weights of the \ac{CC} offline, which notably reduces the computational complexity. 

{\color{reviewer1}
\ac{AIC} and time-domain strategies are usually combined. The method in \cite{Diez19} jointly optimizes the \ac{AIC} and the pulse waveform employed in each carrier. The technique in \cite{Hussain2022} follows a similar approach but using the same pulse in all carriers. Since the optimization is performed offline in both methods, \cite{Hussain2022} has lower complexity than \cite{Diez19} but higher \ac{OOBE}. 

Unfortunately, the discussed methods are impractical for \ac{CR} because their solution has to be recomputed whenever the spectral location of the band where the \ac{OOBE} is to be lowered, hereafter denoted as notched band, changes. }This work addresses the reduction of the \ac{OOBE} of \ac{OFDM} signals for \ac{CR} applications, where spectral resources are sparse, its occupancy changes adaptively and the process has to be transparent to the receiver. 

Two spectral shaping methods are proposed {\color{reviewer2} under the assumption that sequences transmitted in the different carriers are uncorrelated}. One is aimed at reducing the \ac{OOBE} of signals whose passband edges are far enough from each other for their respective spectral shaping problems to be independent, while the other is applicable to arbitrary passband widths. While both are grounded in the generalized pulse given in \cite{Diez19}, the following aspects are innovated. To address the first problem, the notation of the generalized pulses is modified to account for their distance to the passband edge, which is now required to allow them to be adapted to changes in the emission mask, and their optimization is innovated by minimizing their \ac{OOBE} only in the vicinity of the considered edge. To address the second problem, the framework in \cite{Diez19} is innovated in a twofold way: a modified pulse that embeds additional sidelobe reduction terms is defined and a novel optimization method that minimizes the \ac{OOBE} achieved by the considered pulses in a set of signals with different passbands is proposed. Furthermore, simple operations that allow obtaining frequency-shifted and frequency-reversed versions of the pulses employed in both problems are provided. As a result, a set of precomputed pulses can be applied to passbands with different locations and widths by means of these simple transformations. The first method has lower complexity but the second is applicable to passbands of arbitrary width. In summary, the following contributions are made:

\begin{itemize}

\item {It defines a framework that allows the dynamical adaptation of precomputed spectral shaping solutions to changes in the emission mask.}

\item{The adaptation of the precalculated solutions is computationally simple and an efficient \ac{IDFT}-based implementation of the \ac{AST} term is given.} 

\item {It proposes a low-complexity technique that can be applied to lower the \ac{OOBE} both in the sidebands of the transmitted signal and to create multiple notches inside it, irrespective of their spectral location and width. The proposal retains the benefits of the generalized pulse: it is transparent for the receiver; it does not increase the \ac{PAPR} and the \ac{PSD} of the resulting signal can be calculated analytically.}

\end{itemize}

The rest of the paper is organized as follows. Section \ref{Problem_statement} states the problem to be addressed when the passband is wide. The modified notation of the generalized pulses is given along with their optimization procedure. In Section \ref{Frequency_transformation}, the transformations needed to adapt the proposed pulses to different spectral locations are derived. Section \ref{Generalized_Spectral_Shaping} proposes a second and more comprehensive pulse to solve spectral shaping problems with arbitrarily narrow passband, along with the accompanying optimization method. The \ac{OOBE} reduction achieved with the proposed methods is numerically assessed in Section \ref{Numerical_results}. Finally, Section \ref{Conclusion} concludes the work.
\subsection{Notation and Definitions}
Scalar variables are written using italic letters. Matrices and column vectors are written in boldface, the former in capital letters. Sets are denoted using calligraphic letters, e.g. $\mathcal{A}$, and their cardinality as $|\,\cdot\,|$. The Hermitian, the conjugate and transpose operators are denoted as $(\cdot)^{\textrm{H}}$, $(\cdot)^*$ and $(\cdot)^{\textrm{T}}$, respectively. The floor and round functions are denoted as $\lfloor \cdot \rfloor$ and $\lceil \cdot \rfloor$, respectively. The imaginary unit is written as $j = \sqrt{-1}$. 
The operation $a\; \mathrm{mod}\; n$ returns the remainder after division of $a$ by $n$. The relation $a \equiv b\; (\mathrm{mod}\;n)$ denotes that $a$ is congruent to $b$ modulo $n$, meaning that $a \;\mathrm{mod}\; n = b \;\mathrm{mod}\; n$. $\mathbf{I}_M$ is the $M \times M$ identity matrix, while $\mathbf{0}_{M,N}$ is an $M \times N$ zero matrix. An $M \times M$ diagonal matrix with elements $\left\{ x_1, \dots, x_M \right\}$ is denoted as $\textrm{diag}\{ x_1, \dots, x_M \}$. The complex exponential is written as $w_N^{kn} = e^{j\frac{2\pi}{N}kn}$ and the vector $\mathbf{w}_N^k = \left[ w_N^0, \dots, w_N^{k(N-1)} \right]^{\textrm{T}}$.

The considered \ac{OFDM} system has $N$ carriers, which can be classified into three sets according to their functionality: data, \ac{CC} and null. Data carriers are used for conveying information and their indexes are given by the set $\mathcal{D}=\left\{ d_1, \dots, d_{|\mathcal{D}|} \right\}$. The subset of data carriers that employ the proposed pulses is denoted as $\mathcal{D}^{\textrm{h}}$, such that $\mathcal{D}^{\textrm{h}} \subset \mathcal{D}$.  \ac{CC} are exclusively used to shape the spectrum and their indexes are $\mathcal{C}=\left\{ c_1, \dots, c_{|\mathcal{C}|} \right\}$. Null carriers have no allocated power.

\section{Spectral Shaping for Transmission Bands with Wide Bandwidth: the Local Optimization Method}\label{Problem_statement}
\subsection{Problem Statement}
The discrete-time low-pass equivalent expression of an \ac{OFDM} signal can be written as
\begin{equation}
x(n) = \sum^{\infty}_{u=-\infty}x_u(n-uN_{\textrm{s}}),
\label{eq:bgnd:senalOFDM}
\end{equation}
where $N_{\textrm{s}} = N+N_{\textrm{GI}}$ is the symbol period, $N$ is the size of the \ac{DFT} and $N_{\textrm{GI}}$ is the number of samples in the guard interval. The $u$-th \ac{OFDM} symbol is given by
\begin{equation}
x_u(n) = \sum_{k \in \mathcal{D}} p_k(n) s_k(u)
\label{eq:bgnd:simboloOFDMconv}
\end{equation}
where $s_k(u)$ is the modulating symbol transmitted in carrier $k$ and $p_k(n)$ is the basic pulse used in carrier $k$, which is obtained by modulating the shaping pulse $g(n)$,
\begin{equation}
p_k(n) = g(n)w_N^{k(n-N_{\textrm{GI}})},
\label{eq:bgnd:pulsoBase}
\end{equation}
where $g(n)$ can have smooth transitions, as illustrated in Fig. \ref{fig:shapingPulse}, for better spectral confinement. 

\makeatletter \if@twocolumn
\begin{figure}[!tb]
\centering
\includegraphics[width=\columnwidth]{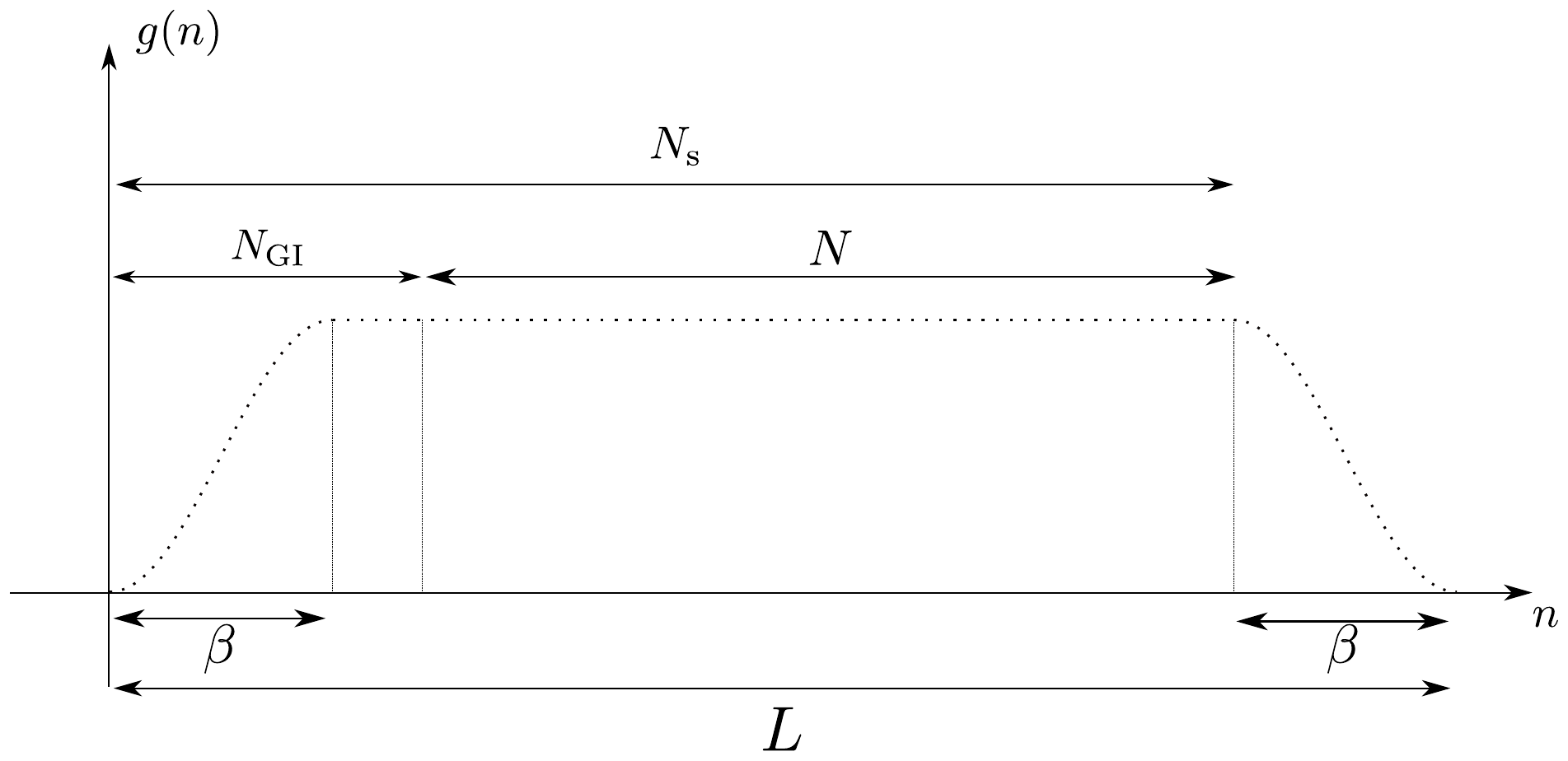}
\caption{Shaping pulse $g(n)$ with smooth transitions and non-zero samples only in the interval $n \in \left\{0, \dots, L-1\right\}$.}
\label{fig:shapingPulse}
\end{figure}
\else
\begin{figure}[!htb]
\centering
\includegraphics[width=8.5cm]{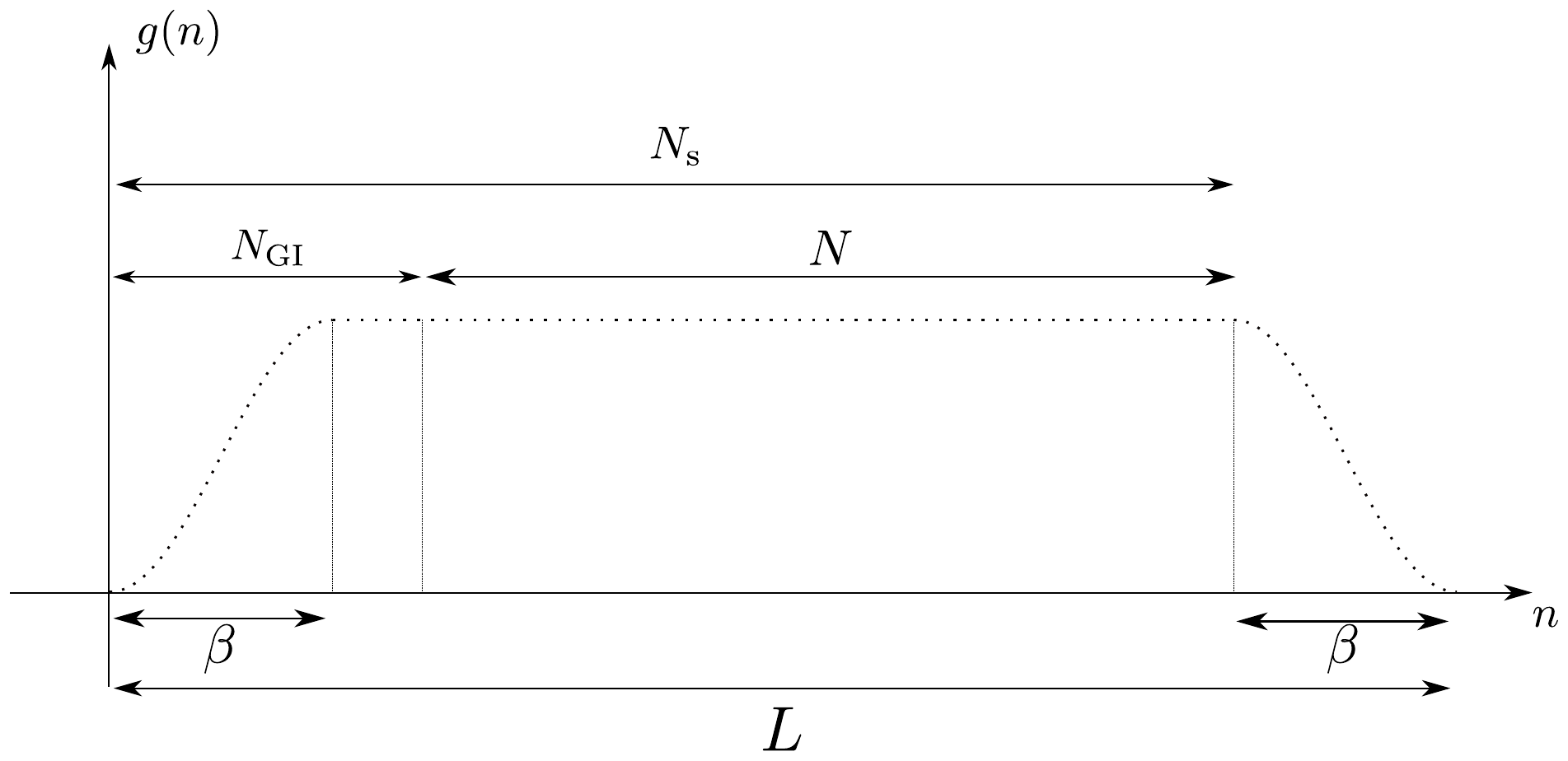}
\vspace*{-0.5cm}
\caption{Shaping pulse $g(n)$ with smooth transitions and non-zero samples only in the interval $n \in \left\{0, \dots, L-1\right\}$.}\label{fig:shapingPulse}
\vspace*{-0.5cm}
\end{figure}
\fi \makeatother
Let us consider the emission mask in Fig. \ref{fig:isolatedPassband}, which corresponds to an isolated passband whose left and right edges are located at carrier indexes $l_{\textrm{l}}$ and $l_{\textrm{r}}$, respectively. These edges determine the final/initial limit of the left/right notched bands, denoted as $\mathcal{B}^+_{\textrm{n}}(l_{\textrm{l}})$ and $\mathcal{B}^-_{\textrm{n}}(l_{\textrm{r}})$, respectively. It is assumed that both notched bands have equal normalized frequency span, denoted as $B_{\textrm{n}}$, which implies that the same objective \ac{OOBE} is imposed in both notched bands. These notched bands extend circularly within the discrete-time spectrum, since the latter is periodic with periodicity $1$. Hence, if the value of $l_{\textrm{l}}$, $l_{\textrm{r}}$ and $B_{\textrm{n}}$ are such that the notched bands extend beyond the $f \in \left(0, 1 \right]$ limits (e.g., after a frequency-shift is applied) the portion of the notched band that exceeds one of the limits appears by the other end of the spectrum.

\makeatletter \if@twocolumn
\begin{figure}[!t]
\centering
\includegraphics[width=8.0cm]{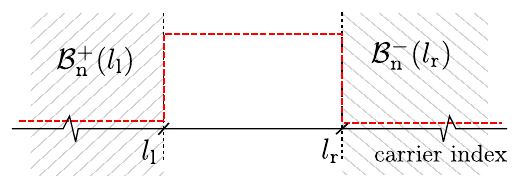}
\caption{Transmitter \ac{PSD} mask with passband left and right edges in $l_\textrm{l}$ and $l_\textrm{r}$ $\mathcal{B}_\textrm{n}^+(l_\textrm{l})$ ans $\mathcal{B}_\textrm{n}^-(l_\textrm{r})$ denote the notched bands.}
\label{fig:isolatedPassband}
\end{figure}
\else
\begin{figure}[!h]
\centering
\includegraphics[width=9.0cm]{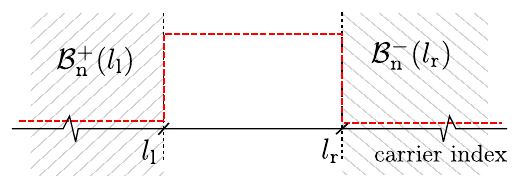}
\vspace*{-0.5cm}
\caption{Transmitter \ac{PSD} mask with passband left and right edges in $l_\textrm{l}$ and $l_\textrm{r}$. $\mathcal{B}_\textrm{n}^+(l_\textrm{l})$ ans $\mathcal{B}_\textrm{n}^-(l_\textrm{r})$ denote the notched bands.}
\label{fig:isolatedPassband}
\vspace{-0.5cm}
\end{figure}

To comply with this mask, the spectral shaping strategy proposed in \cite{Diez19} can be employed. It consists of a pulse, referred to as generalized pulse, which combines \ac{AIC} and \ac{AST}, and is used by the data carriers located closest to the edges of the passbands, since they are the ones that contribute the most to the \ac{OOBE}. The method can be applied in a receiver agnostic way and the generalized pulses can be computed offline. However, an important drawback is that they have to be recomputed every time the emission mask changes. This recomputation is impractical to be performed online in the transmitter as it involves solving a costly minimization problem. Hence, a modification of this method that overcomes this downside is proposed in this work.

Let us denote the proposed pulse used in carrier $k$ as $h_k^{(i)}(n)$, where the index $i$ represents the relative position of the carrier $k$ with respect to the nearest edge of the passband. So, $i > 0$ ($i < 0$) corresponds to carriers near the left (right) edge. This emphasizes that the pulse employed in a given carrier not only depends on the carrier $k$ itself, but also on its relative position with respect to the closest passband edge.
Assuming that the passband in Fig. \ref{fig:isolatedPassband} is wide enough so that the proposed pulses employed at one end of the passband do not interfere with those located at the opposite edge, the spectral shaping problems at both ends can be solved independently. 

Fig. \ref{fig:bandBorderDefinitions} depicts the arrangement of the carriers employed in the spectral shaping solution at each edge of a wide passband. As seen, $h_k^{(i)}(n)$ is used only by the data carriers located closest to the edges of the passband (displayed in blue). The number of data carriers that use $h_k^{(i)}(n)$ at each end of the passband is denoted by $N_{\textrm{h}}$, which is a design parameter that influences the complexity and the performance of the method. Its upper bound, denoted by $N_{\textrm{h\_max}}$, is selected such that using the proposed pulse in one additional carrier yields a negligible \ac{OOBE} reduction. In this section, $N_{\textrm{h}}=N_{\textrm{h\_max}}$ is considered, which allows disregarding the energy emitted by the carriers that do not use the proposed pulse.

The \ac{CC}, in yellow in Fig. \ref{fig:bandBorderDefinitions}, can be located inside the passband (in-band) and out-of-band, being $N_{\textrm{ci}}$ and $N_{\textrm{co}}$, respectively, the number of \ac{CC} located at each side of the edge of the passband. The total number is denoted by $N_{\textrm{CC}} = N_{\textrm{ci}}+N_{\textrm{co}}+1$, as there is one \ac{CC} right at the edge. Finally, the data carriers in black use the basic pulse $p_k(n)$. 

\fi \makeatother
\makeatletter \if@twocolumn
\begin{figure}[!t]
\centering
\includegraphics[width=\columnwidth]{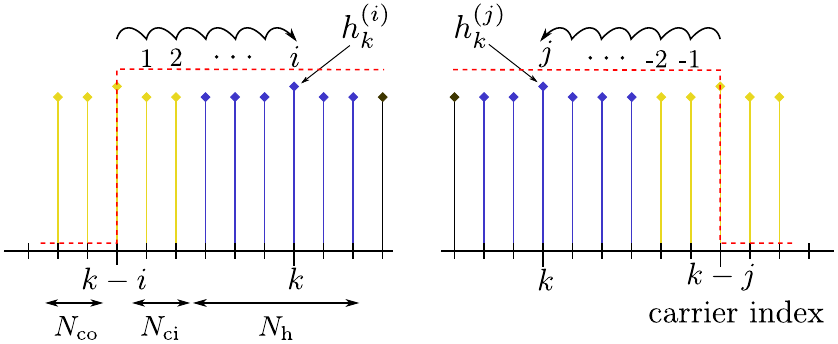}
\caption{Detailed representation of the carriers at the left and right edges of a passband.}
\label{fig:bandBorderDefinitions}
\end{figure}
\else
\begin{figure}[!h]
\centering
\includegraphics[width=9.5cm]{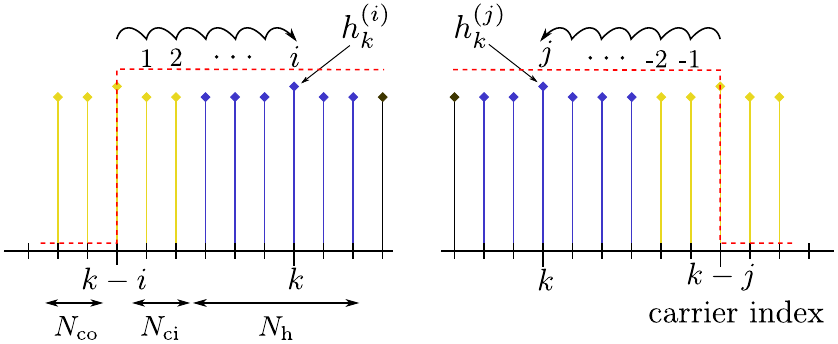}
\vspace*{-0.25cm}
\caption{Detailed representation of the carriers at the left and right edges of a passband.}
\label{fig:bandBorderDefinitions}
\end{figure}
\vspace*{-0.25cm}
\fi \makeatother

When the proposed pulses are employed, the $u$-th \ac{OFDM} symbol is given by
\makeatletter \if@twocolumn
\begin{equation} 
\begin{aligned}
x_u(n) = &\sum_{k=l_{\textrm{l}}+N_{\textrm{ci}}+1}^{l_{\textrm{l}}+N_{\textrm{ci}}+N_{\textrm{h}}} h_k^{(k-l_{\textrm{l}})}(n) s_k(u) +\\ 
&\sum_{k=l_{\textrm{l}}+N_{\textrm{ci}}+N_{\textrm{h}}+1}^{l_{\textrm{r}}-N_{\textrm{ci}}-N_{\textrm{h}}-1} \!\!\!\!\!\!\!\!p_k(n) s_k(u) +\\ 
&\sum_{k=l_{\textrm{r}}-N_{\textrm{ci}}-N_{\textrm{h}}}^{l_{\textrm{r}}-N_{\textrm{ci}}-1} \!\!\!\!\!\!\!\!h_k^{(k-l_{\textrm{r}})}(n) s_k(u),
\end{aligned}
\label{eq:Pgen:simboloOFDMgen}
\end{equation}
\else
\begin{equation}
x_u(n) = \sum_{k=l_{\textrm{l}}+N_{\textrm{ci}}+1}^{l_{\textrm{l}}+N_{\textrm{ci}}+N_{\textrm{h}}} h_k^{(k-l_{\textrm{l}})}(n) s_k(u) + \sum_{k=l_{\textrm{l}}+N_{\textrm{ci}}+N_{\textrm{h}}+1}^{l_{\textrm{r}}-N_{\textrm{ci}}-N_{\textrm{h}}-1} \!\!\!\!\!\!\!\! p_k(n) s_k(u) + \sum_{k=l_{\textrm{r}}-N_{\textrm{ci}}-N_{\textrm{h}}}^{l_{\textrm{r}}-N_{\textrm{ci}}-1} \!\!\!\!\!\!\!\! h_k^{(k-l_{\textrm{r}})}(n) s_k(u).
\label{eq:Pgen:simboloOFDMgen}
\end{equation}
\fi \makeatother

Assuming that the sequence $s_k(u)$ transmitted in each data carrier is white (i.e., flat \ac{PSD} and zero mean), with variance  $\sigma_k^2$, and that sequences transmitted in different carriers are independent\footnote{{\color{reviewer2}Actual systems include a scrambler to yield white sequences, since colored ones degrade the performance of some receiver algorithms, may cause peaks in the \ac{PSD} and increase the \ac{PAPR}. However,  the redundancy introduced by the channel encoder might affect the whiteness. While the work in \cite{Mannerkoski00} shows that many common convolutional and block channel codes (e.g., Reed-Solomon) yield white sequences, the validity of this assumption in the target system should be assessed.}}, the \ac{PSD} of (\ref{eq:bgnd:senalOFDM}) can be analytically computed as 
\makeatletter \if@twocolumn
\begin{equation} 
\begin{aligned}
S(f) = &\frac{1}{N_{\textrm{s}}} \sum_{k=l_{\textrm{l}}+N_{\textrm{ci}}+1}^{l_{\textrm{l}}+N_{\textrm{ci}}+N_{\textrm{h}}} \sigma_k^2 \left| H_k^{(k-l_{\textrm{l}})}(f) \right|^2 +\\
&\frac{1}{N_{\textrm{s}}}\sum_{k=l_{\textrm{l}}+N_{\textrm{ci}}+N_{\textrm{h}}+1}^{l_{\textrm{r}}-N_{\textrm{ci}}-N_{\textrm{h}}-1} \sigma_k^2 \big| P_k(f) \big|^2 +\\
&\frac{1}{N_{\textrm{s}}} \sum_{k=l_{\textrm{r}}-N_{\textrm{ci}}-N_{\textrm{h}}}^{l_{\textrm{r}}-N_{\textrm{ci}}-1} \sigma_k^2 \left| H_k^{(k-l_{\textrm{r}})}(f) \right|^2
\end{aligned}
\label{eq:bgnd:PSDsenalOFDM}
\end{equation}
\else
\begin{equation}
S(f) = \frac{1}{N_{\textrm{s}}} \sum_{k=l_{\textrm{l}}+N_{\textrm{ci}}+1}^{l_{\textrm{l}}+N_{\textrm{ci}}+N_{\textrm{h}}} \!\!\!\!\!\sigma_k^2 \left| H_k^{(k-l_{\textrm{l}})}(f) \right|^2 +
\frac{1}{N_{\textrm{s}}}\sum_{k=l_{\textrm{l}}+N_{\textrm{ci}}+N_{\textrm{h}}+1}^{l_{\textrm{r}}-N_{\textrm{ci}}-N_{\textrm{h}}-1}\!\!\!\!\!\!\!\!\! \sigma_k^2 \big| P_k(f) \big|^2 + \frac{1}{N_{\textrm{s}}} \sum_{k=l_{\textrm{r}}-N_{\textrm{ci}}-N_{\textrm{h}}}^{l_{\textrm{r}}-N_{\textrm{ci}}-1}\!\!\!\!\!\!\! \sigma_k^2 \left| H_k^{(k-l_{\textrm{r}})}(f) \right|^2,
\label{eq:bgnd:PSDsenalOFDM}
\end{equation}
\fi \makeatother
where $P_k(f)$ and $H_k^{(i)}(f)$ are the Fourier transforms of $p_k(n)$ and $h_k^{(i)}(n)$, respectively.  Given the vector form of these pulses, $\mathbf{p}_k = \left[ p_k(0), \dots,\right.$ $\left. p_k(L-1) \right]^{\textrm{T}}$ and $\mathbf{h}_k^{(i)} = \left[ h_k^{(i)}(0), \dots, h_k^{(i)}(L-1) \right]^{\textrm{T}}$, these Fourier transforms can be compactly written as,
\begin{equation}
P_k(f) = \mathbf{f}_L^{\textrm{H}}(f) \mathbf{p}_k, \;\;\;\;\;\;\;\;\;\; H_k^{(i)}(f) = \mathbf{f}_L^{\textrm{H}}(f) \mathbf{h}_k^{(i)},
\label{eq:bgnd:TFpulsos}
\end{equation}
where $\mathbf{f}_L^{\textrm{H}}(f) = \left[ 1, e^{-j2\pi f}, \dots, e^{-j2\pi f(L-1)} \right]$.



When a wide passband is assumed and $N_{\textrm{h}}=N_{\textrm{h\_max}}$, the \ac{OOBE} is dominated by the energy emitted by the $N_{\textrm{h}}$ pulses located closest to the notched band. Under this assumption, the power of the \ac{OFDM} signal in the notched bands can be expressed as,
\begin{equation}
\makeatletter
\if@twocolumn
    \begin{aligned}
        \textrm{OOBE} = &\int_{\mathcal{B}_{\textrm{n}}^+(l_{\textrm{l}}) \cup \mathcal{B}_{\textrm{n}}^-(l_{\textrm{r}})} \!\!\!\!\!\!\!\!\!\!\!\!  S(f) df \approx \frac{1}{N_{\textrm{s}}} \sum_{k=l_{\textrm{l}}+N_{\textrm{ci}}+1}^{l_{\textrm{l}}+N_{\textrm{ci}}+N_{\textrm{h}}} \sigma_k^2 E_{k}^{(k-l_{\textrm{l}})} +\\ &\frac{1}{N_{\textrm{s}}} \sum_{k=l_{\textrm{r}}-N_{\textrm{ci}}-N_{\textrm{h}}}^{l_{\textrm{r}}-N_{\textrm{ci}}-1} \sigma_k^2 E_{k}^{(k-l_{\textrm{r}})}
    \end{aligned}
\else
    \begin{aligned}
        \textrm{OOBE} = &\int_{\mathcal{B}_{\textrm{n}}^+(l_{\textrm{l}}) \cup \mathcal{B}_{\textrm{n}}^-(l_{\textrm{r}})} \!\!\!\!\!\!\!\!\!\!\!\!  S(f) df \approx \frac{1}{N_{\textrm{s}}} \sum_{k=l_{\textrm{l}}+N_{\textrm{ci}}+1}^{l_{\textrm{l}}+N_{\textrm{ci}}+N_{\textrm{h}}} \sigma_k^2 E_{k}^{(k-l_{\textrm{l}})} + \frac{1}{N_{\textrm{s}}} \sum_{k=l_{\textrm{r}}-N_{\textrm{ci}}-N_{\textrm{h}}}^{l_{\textrm{r}}-N_{\textrm{ci}}-1} \sigma_k^2 E_{k}^{(k-l_{\textrm{r}})}
    \end{aligned}
\fi
\makeatother
\label{eq:bgnd:potenciaOFDM}
\end{equation}
where
\begin{equation}
\begin{aligned}
E_k^{(i)} = \left\{ \begin{aligned}
    &\int_{\mathcal{B}_{\textrm{n}}^+(k-i)} \left| H_k^{(i)}(f) \right|^2 df,  \quad  \textrm{for } i > 0\\
    &\int_{\mathcal{B}_{\textrm{n}}^-(k-i)} \left| H_k^{(i)}(f) \right|^2 df,  \quad  \textrm{for } i < 0\\
\end{aligned} \right.
\end{aligned},
\label{eq:Pgen:energiaPulsoGen}
\end{equation}
denotes the energy emitted by $h_k^{(i)}(n)$ to the closest notched band. Expression (\ref{eq:bgnd:potenciaOFDM}) consists of two separate summation terms, corresponding to the carriers located at each end of the passband. Since there is no interference between them, the spectral shaping of the left and right edges of the passband can be performed separately. Moreover, the \ac{OOBE} is reduced by minimizing (\ref{eq:Pgen:energiaPulsoGen}) for each carrier with pulse $h_k^{(i)}(n)$, as emissions produced by different carriers are independent.
\subsection{Pulse Definition}
\label{The_generalized_pulse}
The expression of the  proposed pulse used at the left edge of a wide passband is
\begin{equation}
\mathbf{h}_k^{(i)} = \mathbf{p}_k + \mathbf{C}_{k-i}^{+} \boldsymbol{\alpha}_k^{(i)} + \mathbf{t}_k^{(i)}, \quad i>0.
\label{eq:Pgen:pGenFlIzq}
\end{equation} 
The \ac{RHS} is the $L \times 1$ vector form of the basic pulse in (\ref{eq:bgnd:pulsoBase}) plus the \ac{AIC} and \ac{AST} terms used to reduce its \ac{OOBE}. 
$\mathbf{C}_{k-i}^+$ is an $L \times N_{\textrm{CC}}$ matrix comprising a set of pulses,
\begin{equation}
\begin{aligned}
\mathbf{C}_l^{+} &= \left[ \mathbf{p}_{l-N_{\textrm{co}}}, \dots, \mathbf{p}_l, \dots, \mathbf{p}_{l+N_{\textrm{ci}}} \right],\\
\end{aligned}
\label{eq:Pgen:matrizCCizq}
\end{equation}
which are linearly combined with the complex coefficients $\boldsymbol{\alpha}_k^{(i)} = \left[ \alpha_k^{(i)}(1), \dots, \alpha_k^{(i)}(N_{\textrm{CC}}) \right]^{\textrm{T}}$. These are the \ac{CC}, which are distributed around the left edge of the passband, located at $k-i$, as they have better control over the energy emissions in the notched band. Also, note that when $k$ rises/decreases $i$ rises/decreases too, keeping $(k-i)$ constant. Consequently, all proposed pulses used at the given edge will make use of the same set of \ac{CC}. 

Finally, $\mathbf{t}_k^{(i)}$ is a vector of $L$ samples used to conform the time-domain shape of the initial and final boundaries of the pulse, and is referred to as transition pulse. Since only its first and last $\beta$ samples are nonzero, it can be expressed as
\begin{equation}
\mathbf{t}_k^{(i)} = \left[ \begin{array}{c}
\mathbf{I}_{\beta} \quad\quad \mathbf{0}_{\beta,\beta}\\ \mathbf{0}_{L-2\beta,2\beta} \\\mathbf{0}_{\beta,\beta} \quad\quad \mathbf{I}_{\beta}
\end{array} \right] \boldsymbol{\zeta}_k^{(i)},
\label{eq:Pgen:pulsTransicZeta}
\end{equation}
where $\boldsymbol{\zeta}_k^{(i)} = \left[ \zeta_k^{(i)}(0), \dots, \zeta_k^{(i)}(2\beta -1) \right]^{\textrm{T}}$. Both $\boldsymbol{\alpha}_k^{(i)}$ and $\boldsymbol{\zeta}_k^{(i)}$ are jointly optimized to minimize the energy emitted by the proposed pulse to the notched band. This energy emission is given by (\ref{eq:Pgen:energiaPulsoGen}) and can be expressed in matrix form as
\begin{equation}
E_k^{(i)} = (\mathbf{h}_k^{(i)})^{\textrm{H}} \boldsymbol{\Phi}_{\mathcal{B}^+_{\textrm{n}}(l_{\textrm{l}})}\mathbf{h}_k^{(i)}, \quad i > 0,
\label{eq:Pgen:energiaPulsGenbis}
\end{equation}
where $\boldsymbol{\Phi}_{\mathcal{B}^+_{\textrm{n}}(l_{\textrm{l}})} = \int_{\mathcal{B}^+_{\textrm{n}}(l_{\textrm{l}})} \mathbf{f}_L(f) \mathbf{f}_L^{\textrm{H}}(f) df$ is an $L \times L$ Hermitian Toeplitz matrix that depends only on the considered frequency range. 

The expression for the proposed pulses near the right edges, $\mathbf{h}_k^{(i)}$ with $i<0$, is similar to that in (\ref{eq:Pgen:pGenFlIzq}) only that matrix $\mathbf{C}_{l}^+$ is substituted for $\mathbf{C}_l^{-} = \left[ \mathbf{p}_{l+N_{\textrm{co}}}, \dots, \mathbf{p}_l, \dots, \mathbf{p}_{l-N_{\textrm{ci}}} \right].$
The rest of definitions are analogous to those presented for the left edge. Hence, the expressions associated to the left edge will be hereafter presented first and, for the sake of conciseness, those associated to the right edge will appear only when needed.  

The pulse $h_k^{(i)}(n)$ includes the \ac{AIC} and \ac{AST} terms to cancel the \ac{OOBE} of the basic pulse $p_k(n)$, just as the generalized pulse in \cite{Diez19} does. However, the distinctive feature of $h_k^{(i)}(n)$ is that the relative position of the carrier $k$ with respect to the edge of the passband, represented by the index $i$, is explicitly stated. When the \ac{AIC} and \ac{AST} terms are determined for a given emission mask, the distance to the passband edges is implicitly included in the carrier index. However, when these precomputed terms are to be employed in a frequency-shifted version of the emission mask, their relative distance to the new passband edge must be the same as in the original one. To this end, the distance of the carrier index to the passband edge for which they were optimized must be stated as part of the pulse definition. 

An additional innovation of the proposed method with respect to the framework in \cite{Diez19} is the cost function in (\ref{eq:Pgen:energiaPulsoGen}), which only considers the \ac{OOBE} in the closest notched band. This is a key aspect, since it makes the optimization problem independent of the actual passband width. This allows reusing the \ac{AIC} and \ac{AST} terms computed for a given emission mask in other scenarios, as long as their passband is wide enough for the proposed pulses employed at one end of the band not to interfere with those located at the opposite edge.

\subsection{Local Optimization of the Proposed Pulses}
The optimal coefficients of $h_k^{(i)}(n)$ in carrier $k$ are obtained by minimizing (\ref{eq:Pgen:energiaPulsGenbis}) as
\begin{equation}
\widehat{\boldsymbol{\gamma}}_k^{(i)} = \left[ \begin{array}{c}
\widehat{\boldsymbol{\alpha}}_k^{(i)}\\ \widehat{\boldsymbol{\zeta}}_k^{(i)}
\end{array} \right] = \argmin_{\boldsymbol{\alpha}_k^{(i)}, \boldsymbol{\zeta}_k^{(i)}} \left\{ E_k^{(i)} \right\}.
\label{eq:Pgen:argminEnergia}
\end{equation}

The unconstrained minimization in (\ref{eq:Pgen:argminEnergia}) may cause undesirable \ac{PSD} peaks in the passband. This can be avoided by constraining the absolute value of the real and imaginary parts of $\boldsymbol{\alpha}_k^{(i)}$ and $\boldsymbol{\zeta}_k^{(i)}$.

Since (\ref{eq:Pgen:argminEnergia}) is data independent, it can be precomputed offline and applied online to the transmitted signal. Besides, neither the \ac{CC} nor the transition pulses interfere with the regular reception of the signal, because the latter are discarded along with the guard interval and the \ac{CC} are orthogonal to the data carriers. Hence, a conventional \ac{OFDM} receiver can be employed.
\subsection{Transition Pulses with Harmonically Designed Boundaries}
The transition pulses can be harmonically designed by expressing their non-zero samples using a $\beta$-samples \ac{IDFT} \cite{Diez19}. Since the \ac{OOBE} reduction is mainly due to the \ac{IDFT} coefficients corresponding to frequencies closer to the notched band, distant ones can be left out of the \ac{IDFT} matrix. This substantially reduces both the number of coefficients to be computed in the optimization procedure and the complexity of its real-time implementation. Let us define the $\beta \times N_{\textrm{QQ}}$ matrices, $\mathbf{Q}_l^{+}$ and $\mathbf{Q}_l^{-}$, which contain the $\beta$-\ac{IDFT} columns corresponding to the $N_{\textrm{QQ}}$ frequencies situated closer to index $l$, which represents the position of the corresponding left/right edge of the passband. The expression for the former is
\begin{equation}
\mathbf{Q}_l^+ \!= \!\! \left\{
\begin{aligned}
\!\!&\left[ \mathbf{w}_{\beta}^{\frac{l}{R}-N_{\textrm{q}}}\!\!\!\!, \dots, \mathbf{w}_{\beta}^{\frac{l}{R}}, \dots, \mathbf{w}_{\beta}^{\frac{l}{R}+N_{\textrm{q}}} \right]  \textrm{when $\frac{l}{R} \in \mathbb{Z}$}\\
\!\!&\left[ \mathbf{w}_{\beta}^{\lfloor\frac{l}{R}\rfloor-N_{\textrm{q}}+1}\!\!\!\!\!\!, \dots, \mathbf{w}_{\beta}^{\lfloor\frac{l}{R}\rfloor}, \dots, \mathbf{w}_{\beta}^{\lfloor\frac{l}{R}\rfloor+N_{\textrm{q}}} \right]  \textrm{otherwise}
\end{aligned},
\right.
\label{eq:Pgen:matricesQl1}
\end{equation}
where $R = \frac{N}{\beta} \in \mathbb{Q}^+$ denotes the relation between the number of carriers in the \ac{OFDM} symbols, $N$, and the number of non-zero samples at each side of the transition pulses, $\beta$. 

Fig. \ref{fig:QmatDefinition} depicts the distribution of the frequency terms in $\mathbf{Q}_l^+$ and $\mathbf{Q}_l^-$ (in green) with respect to the edge of the passband: $N_{\textrm{q}}$ are employed at each side of such edge and the circled numbers denote the matrix column corresponding to each of them. Notice that when $\frac{l}{R} \in \mathbb{Z}$ an additional frequency term is employed, located right on the edge of the passband, besides the $2N_{\textrm{q}}$ terms that are symmetrically distributed around it, being $N_{\textrm{QQ}}=2N_{\textrm{q}}+1$. When $\frac{l}{R} \notin \mathbb{Z}$, the relative location of the $N_{\textrm{QQ}}=2N_{\textrm{q}}$ \ac{IDFT} terms with respect to the passband edge depends on the position of the notched band with respect to the edge  (left/right) and the value of $l \textrm{ mod }R$.

\makeatletter \if@twocolumn
\begin{figure}[!b]
\centering
\vspace{-0.5cm}
\includegraphics[width=\columnwidth]{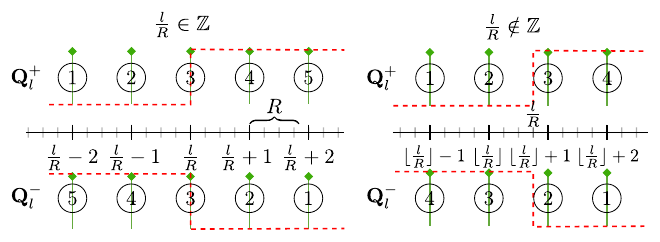}
\caption{Location of the frequency terms contained in matrices $\mathbf{Q}_l^{+}$ and $\mathbf{Q}_l^{-}$ with respect to the passband edge when $N_{\textrm{q}} = 2$ and $R=4$.}
\label{fig:QmatDefinition}
\end{figure}
\else
\begin{figure}[!h]
\centering
\includegraphics[width=10.5cm]{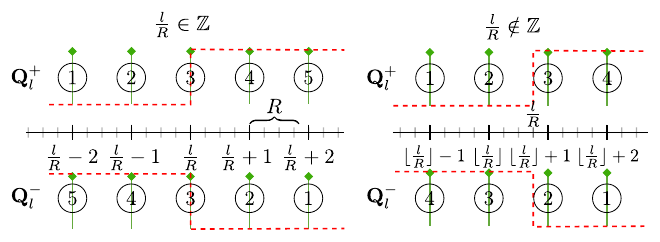}
\vspace{-0.5cm}
\caption{Location of the frequency terms contained in $\mathbf{Q}_l^{+}$ and $\mathbf{Q}_l^{-}$ with respect to the passband edge when $N_{\textrm{q}} = 2$ and $R=4$.}
\label{fig:QmatDefinition}
\vspace{-0.75cm}
\end{figure}
\fi \makeatother

For the proposed pulses located at the right edge, the expression of $\mathbf{Q}_l^{-}$ is
\begin{equation}
\mathbf{Q}_l^- \!= \!\!\left\{
\begin{aligned}
\!\!&\left[ \mathbf{w}_{\beta}^{\frac{l}{R}+N_{\textrm{q}}}\!\!\!\!, \dots, \mathbf{w}_{\beta}^{\frac{l}{R}}, \dots, \mathbf{w}_{\beta}^{\frac{l}{R}-N_{\textrm{q}}} \right]  \textrm{when $\frac{l}{R} \in \mathbb{Z}$},\\
\!\!&\left[ \mathbf{w}_{\beta}^{\lfloor\frac{l}{R}\rfloor+N_{\textrm{q}}}\!\!\!\!\!, \dots, \mathbf{w}_{\beta}^{\lfloor\frac{l}{R}\rfloor}, \dots, \mathbf{w}_{\beta}^{\lfloor\frac{l}{R}\rfloor-N_{\textrm{q}}+1} \right] \textrm{otherwise}.
\end{aligned} \right.
\label{eq:Pgen:matricesQr1}
\end{equation}

These matrices are used for the harmonic design of the transition pulses, whose new expression for the proposed pulses located at the left edge of the passband is
\begin{equation}
\mathbf{t}_k^{(i)} = \left[ \begin{array}{c}
\mathbf{Q}_{k-i}^{+} \\ \mathbf{0}_{L-\beta,N_{\textrm{QQ}}}
\end{array} \right] \boldsymbol{\varepsilon}_k^{(i),\textrm{s}} + \left[ \begin{array}{c}
\mathbf{0}_{L-\beta,N_{\textrm{QQ}}} \\ \mathbf{Q}_{k-i}^{+}
\end{array} \right] \boldsymbol{\varepsilon}_k^{(i),\textrm{e}},
\label{eq:Pgen:pulsTransHizq}
\end{equation}
where vectors $\boldsymbol{\varepsilon}_k^{(i),\textrm{s}}$ and $\boldsymbol{\varepsilon}_k^{(i),\textrm{e}}$ contain the corresponding coefficients for the starting and ending transition pulses, respectively. For the pulses located at the right edge, the expression for their transition pulses is like that in (\ref{eq:Pgen:pulsTransHizq}) but with $i<0$ and where $\mathbf{Q}_l^+$ is substituted for $\mathbf{Q}_l^-$.

Finally, when transition pulses are harmonically designed expression (\ref{eq:Pgen:argminEnergia}) can be rewritten as
\begin{equation}
\widehat{\boldsymbol{\gamma}}_k^{(i),\textrm{h}} = \left[ \begin{array}{c}
\widehat{\boldsymbol{\alpha}}_k^{(i)}\\ \widehat{\boldsymbol{\varepsilon}}_k^{(i),\textrm{s}}\\ \widehat{\boldsymbol{\varepsilon}}_k^{(i),\textrm{e}}
\end{array} \right] = \argmin_{\boldsymbol{\alpha}_k^{(i)}, \boldsymbol{\varepsilon}_k^{(i),\textrm{s}}, \boldsymbol{\varepsilon}_k^{(i),\textrm{e}}} \left\{ E_k^{(i)} \right\}.
\label{eq:Pgen:argminEnergiaHarm}
\end{equation}

This minimization should also include constraints for the real and imaginary parts of $\boldsymbol{\alpha}_k^{(i)}$, $\boldsymbol{\varepsilon}_k^{(i),s}$ and $\boldsymbol{\varepsilon}_k^{(i),e}$ to avoid \ac{PSD} peaks in the passband.

{\color{black}\section{Frequency Transformation of the Pulses}
\label{Frequency_transformation}}
The study made so far has presented a method to design the proposed pulses closest to the left edge of a passband, under the assumption that the right edge is far enough for the \ac{OOBE} of these pulses to be negligible in the opposite notched band. This section describes simple transformations that allow the proposed pulses that have been designed for a given passband edge to be applied to a passband edge located at a different frequency and with a different relative position with respect to the notched band (by its right/left).

The energy that the proposed pulse $h_k^{(i)}(n)$, $i>0$, emits to the notched band $\mathcal{B}^+_{\textrm{n}}(k-i)$ is

\makeatletter \if@twocolumn
\begin{equation} 
\begin{aligned}
E_k^{(i)} = \left(\mathbf{p}_k + \mathbf{C}_{k-i}^+ \boldsymbol{\alpha}_k^{(i)} + \mathbf{t}_k^{(i)}\right)^{\textrm{H}} \boldsymbol{\Phi}_{\mathcal{B}^+_{\textrm{n}}(k-i)}\\ \left(\mathbf{p}_k + \mathbf{C}_{k-i}^+ \boldsymbol{\alpha}_k^{(i)} + \mathbf{t}_k^{(i)}\right).
\end{aligned}
\label{eq:localOpt:energiaPulsok}
\end{equation}
\else
\begin{equation} 
E_k^{(i)} = \left(\mathbf{p}_k + \mathbf{C}_{k-i}^+ \boldsymbol{\alpha}_k^{(i)} + \mathbf{t}_k^{(i)}\right)^{\textrm{H}} \boldsymbol{\Phi}_{\mathcal{B}^+_{\textrm{n}}(k-i)} \left(\mathbf{p}_k + \mathbf{C}_{k-i}^+ \boldsymbol{\alpha}_k^{(i)} + \mathbf{t}_k^{(i)}\right).
\label{eq:localOpt:energiaPulsok}
\end{equation}
\fi \makeatother

The complex coefficients $\boldsymbol{\alpha}_k^{(i)}$ and $\mathbf{t}_k^{(i)}$ are locally optimized to reduce the \ac{OOBE} in that notched band. The goal now is to find a relation between these coefficients and those obtained in a similar scenario when the passband  and the notched bands are shifted $\Delta k$ carriers. The energy emitted by $h_{k+\Delta k}^{(i)}(n)$ in the corresponding notched band, $\mathcal{B}^+_{\textrm{n}}(k+\Delta k-i)$, is given by
\makeatletter \if@twocolumn
\begin{equation} 
\begin{aligned}
&E_{k+\Delta k}^{(i)} = \Big( \mathbf{p}_{k+\Delta k} + \mathbf{C}_{k+\Delta k-i}^+ \boldsymbol{\alpha}_{k+\Delta k}^{(i)} + \mathbf{t}_{k+\Delta k}^{(i)} \Big)^{\textrm{H}}\\ &\boldsymbol{\Phi}_{\mathcal{B}^+_{\textrm{n}}(k+\Delta k-i,)} \Big( \mathbf{p}_{k+\Delta k} + \\ &\mathbf{C}_{k+\Delta k-i}^+ \boldsymbol{\alpha}_{k+\Delta k}^{(i)} +  \mathbf{t}_{k+\Delta k}^{(i)} \Big).
\end{aligned}
\label{eq:localOpt:energiaPulsoDesplaz}
\end{equation}
\else
\begin{equation} 
E_{k+\Delta k}^{(i)} = \Big( \mathbf{p}_{k+\Delta k} + \mathbf{C}_{k+\Delta k-i}^+ \boldsymbol{\alpha}_{k+\Delta k}^{(i)} + \mathbf{t}_{k+\Delta k}^{(i)} \Big)^{\textrm{H}} \boldsymbol{\Phi}_{\mathcal{B}^+_{\textrm{n}}(k+\Delta k-i)} \Big( \mathbf{p}_{k+\Delta k} + \mathbf{C}_{k+\Delta k-i}^+ \boldsymbol{\alpha}_{k+\Delta k}^{(i)} + \mathbf{t}_{k+\Delta k}^{(i)} \Big).
\label{eq:localOpt:energiaPulsoDesplaz}
\end{equation}
\fi \makeatother


Considering the following identities (see Appendix A for a proof)
\begin{equation}
\mathbf{p}_{k+\Delta k} = \boldsymbol{\Omega}_{\Delta k} \mathbf{p}_k, \;\;\;\;\;
\mathbf{C}_{k+\Delta k-i}^+ = \boldsymbol{\Omega}_{\Delta k} \mathbf{C}_{k-i}^+,\;\;\;\;\;
\boldsymbol{\Phi}_{\mathcal{B}^+_{\textrm{n}}(k+\Delta k-i)} = \boldsymbol{\Omega}_{\Delta k} \boldsymbol{\Phi}_{\mathcal{B}^+_{\textrm{n}}(k-i)} \boldsymbol{\Omega}_{\Delta k}^{\textrm{H}},
\label{eq:localOpt:identidadesDesplazamiento}
\end{equation}
where $\boldsymbol{\Omega}_{\Delta k} = \textrm{diag} \left\{ w_N^{\Delta k (0-N_{\textrm{GI}})}, \dots , w_N^{\Delta k (L-1-N_{\textrm{GI}})} \right\}$, expression (\ref{eq:localOpt:energiaPulsoDesplaz}) can be transformed into
\begin{equation} 
\makeatletter \if@twocolumn
\begin{aligned}
&E_{k+\Delta k}^{(i)} = \left( \mathbf{p}_k + \mathbf{C}_{k-i}^+\boldsymbol{\alpha}_{k+\Delta k}^{(i)} + \boldsymbol{\Omega}_{\Delta k}^{-1} \mathbf{t}_{k+\Delta k}^{(i)} \right)^{\textrm{H}}\\ &\boldsymbol{\Phi}_{\mathcal{B}^+_{\textrm{n}}(k-i)} \left( \mathbf{p}_k + \mathbf{C}_{k-i}^+\boldsymbol{\alpha}_{k+\Delta k}^{(i)} + \boldsymbol{\Omega}_{\Delta k}^{-1} \mathbf{t}_{k+\Delta k}^{(i)} \right).
\end{aligned}
\else
\begin{aligned}
E_{k+\Delta k}^{(i)} = \left( \mathbf{p}_k + \mathbf{C}_{k-i}^+\boldsymbol{\alpha}_{k+\Delta k}^{(i)} + \boldsymbol{\Omega}_{\Delta k}^{-1} \mathbf{t}_{k+\Delta k}^{(i)} \right)^{\textrm{H}} &\boldsymbol{\Phi}_{\mathcal{B}^+_{\textrm{n}}(k-i)} \left( \mathbf{p}_k + \mathbf{C}_{k-i}^+\boldsymbol{\alpha}_{k+\Delta k}^{(i)} + \boldsymbol{\Omega}_{\Delta k}^{-1} \mathbf{t}_{k+\Delta k}^{(i)} \right).
\end{aligned}
\fi \makeatother
\label{eq:localOpt:energiaPulsoDesplazbis}
\end{equation}

As the magnitude of (\ref{eq:localOpt:energiaPulsok}) and (\ref{eq:localOpt:energiaPulsoDesplazbis}) must be equal, the following relations are obtained
\begin{equation}
\boxed{
\boldsymbol{\alpha}_{k+\Delta k}^{(i)} = \boldsymbol{\alpha}_k^{(i)}, \quad\quad \mathbf{t}_{k+\Delta k}^{(i)} = \boldsymbol{\Omega}_{\Delta k}\mathbf{t}_{k}^{(i)}.}
\label{eq:localOpt:transformacionesDesplazamiento}
\end{equation}

The previous transformation shows the relation between the optimal coefficients associated to proposed pulses located at the same side of the passband edge, for either a right or left edge. In order to determine a relation between the coefficients obtained for two opposite edges, let $h_k^{(i)}(n)$, $i>0$, be a proposed pulse designed to reduce the \ac{OOBE} in the notched band $\mathcal{B}^+_{\textrm{n}}(k-i)$. Given the symmetry properties of the Fourier transform, $\mathbf{h}_k^{(i)*}$ equals the proposed pulse used by carrier $N-k$ to minimize the \ac{OOBE} in $\mathcal{B}^-_\textrm{n}(N-k+i)$, as illustrated in Fig. \ref{fig:freqReversal}. It can be easily proven that $\mathbf{h}_{N-k}^{(-i)}=\mathbf{h}_k^{(i)*}$ is satisfied given the following relations
\begin{equation}
\boldsymbol{\alpha}_{N-k}^{(-i)} = \left(\boldsymbol{\alpha}_k^{(i)}\right)^*,\quad\quad \mathbf{t}_{N-k}^{(-i)} = \left(\mathbf{t}_{k}^{(i)}\right)^*,
\label{eq:localOpt:transformacionesGiroInic}
\end{equation}
and, after applying (\ref{eq:localOpt:transformacionesDesplazamiento}) with $\Delta k = 2k-N$, the following relations are obtained,
\begin{equation}
\makeatletter
\if@twocolumn
\boxed{
\begin{aligned}
&\boldsymbol{\alpha}_{k}^{(-i)} = \left(\boldsymbol{\alpha}_k^{(i)}\right)^*,\\ &\mathbf{t}_k^{(-i)} = \boldsymbol{\Omega}_{2k-N} \mathbf{t}_{N-k}^{(-i)} = \boldsymbol{\Omega}_{2k-N}\left(\mathbf{t}_k^{(i)} \right)^*.
\end{aligned} }
\else
\boxed{
\begin{aligned}
&\boldsymbol{\alpha}_{k}^{(-i)} = \left(\boldsymbol{\alpha}_k^{(i)}\right)^*,\quad &\mathbf{t}_k^{(-i)} = \boldsymbol{\Omega}_{2k-N} \mathbf{t}_{N-k}^{(-i)} = \boldsymbol{\Omega}_{2k-N}\left(\mathbf{t}_k^{(i)} \right)^*.
\end{aligned} }
\fi
\makeatother
\label{eq:localOpt:transformacionesGiro}
\end{equation}

\makeatletter \if@twocolumn
\begin{figure}[!b]
\vspace{-0.75cm}
\centering
\includegraphics[width=\columnwidth]{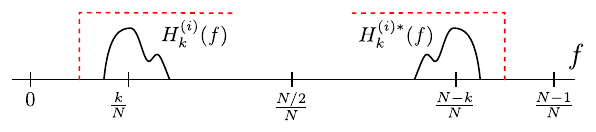}
\caption{Fourier Transform of $\mathbf{h}_k^{(i)*}$ and its mirrored counterpart.}
\label{fig:freqReversal}
\vspace{-0.3cm}
\end{figure}
\else
\begin{figure}[!b]
\vspace{-0.75cm}
\centering
\includegraphics[width=12cm]{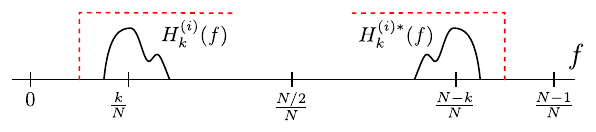}
\vspace{-0.75cm}
\caption{Fourier Transform of $\mathbf{h}_k^{(i)*}$ and its mirrored counterpart.}
\label{fig:freqReversal}
\end{figure}
\fi \makeatother

As a conclusion, (\ref{eq:localOpt:transformacionesDesplazamiento}) and (\ref{eq:localOpt:transformacionesGiro}) allow obtaining the proposed pulses to be applied to the left and right edges of a given passband by frequency-shifting and frequency-reversing the precomputed coefficients obtained through the optimization procedure in (\ref{eq:Pgen:argminEnergia}). These transformations are applied to proposed pulses that occupy the same position within the passband with respect to its edge, given by the index $i$. Since pulses that belong to the same edge will be typically transformed and applied together to a target passband, it is convenient to arrange the vectors of coefficients $\boldsymbol{\gamma}_{k}^{(i)}$, defined in (\ref{eq:Pgen:argminEnergia}), associated to the proposed pulses in a passband left edge located in carrier $l$ in the following $(N_{\textrm{CC}}+2\beta) \times N_{\textrm{h}}$ matrix
\begin{equation}
\boldsymbol{\Gamma}_l^+ = \left[ \boldsymbol{\gamma}_{l+N_{\textrm{ci}}+1}^{(N_{\textrm{ci}}+1)}, \cdots, \boldsymbol{\gamma}_{l+N_{\textrm{ci}}+N_{\textrm{h}}}^{(N_{\textrm{ci}}+N_{\textrm{h}})} \right].
\label{eq:localOptHarm:definicionMatrizGamma}
\end{equation}

Once (\ref{eq:localOptHarm:definicionMatrizGamma}) is computed, the matrix corresponding to any other passband edge location or orientation (left/right) can be easily obtained by virtue of the presented transformations. For instance, the matrix of coefficients associated to the proposed pulses in a right edge located at $l$, which can be written as 
\begin{equation}
\boldsymbol{\Gamma}_l^- = \left[ \boldsymbol{\gamma}_{l-N_{\textrm{ci}}-1}^{(-N_{\textrm{ci}}-1)}, \cdots, \boldsymbol{\gamma}_{l-N_{\textrm{ci}}-N_{\textrm{h}}}^{(-N_{\textrm{ci}}-N_{\textrm{h}})} \right],
\label{eq:localOptHarm:definicionMatrizGammaDcha}
\end{equation}
can be obtained from the coefficients in (\ref{eq:localOptHarm:definicionMatrizGamma}) by means of (\ref{eq:localOpt:transformacionesGiro}). 

\subsection{Frequency Transformations for Transition Pulses with Harmonically Designed Boundaries}
This section presents the frequency-shift and frequency-reversal transformations to be applied to the transition pulses with harmonically designed boundaries. Since changes only affect the part of the identities in (\ref{eq:localOpt:transformacionesDesplazamiento}) and (\ref{eq:localOpt:transformacionesGiro}) relative to the coefficients of the transition pulses, only the transformations concerning $\boldsymbol{\varepsilon}_{k}^{(i),\textrm{s}/\textrm{e}}$ in (\ref{eq:Pgen:pulsTransHizq}) will be presented. 

It must be highlighted that the transformations are defined only for $R = \frac{N}{\beta} \in \mathbb{Z}$ and $\frac{\Delta k}{R} \in \mathbb{Z}$. Fig. \ref{fig:QmatDefinition} illustrates the rationale for this constraint. The frequency terms employed to synthesize the harmonically designed transition pulses (in green) are equidistantly distributed along the transmission band. When $R \in \mathbb{Z}$, these terms coincide with $\beta$ of the $N$-\ac{DFT} carriers (spaced $R$ carriers apart). Hence, these frequency terms are distributed equivalently in all passband edges located $mR$ carriers apart from the original one, with $m \in \mathbb{Z}$. Consequently, only transformations that involve frequency shifts of $\Delta k = mR$ carriers result in a set of coefficients that yields the same \ac{OOBE} reduction as the original ones. 

Denoting by $\boldsymbol{\varepsilon}_k^{(i),s/e}$, $i>0$, the coefficients of the harmonically designed transition pulse in carrier $k$ that minimize the \ac{OOBE} in the notched band $\mathcal{B}_\textrm{n}^+(k-i)$, it can be proved that the coefficients corresponding to the pulse in carrier $k + \Delta k$ that minimize the \ac{OOBE} in the notched band $\mathcal{B}_\textrm{n}^+(k+\Delta k-i)$ can be obtained as
\begin{equation} 
\boxed{
\begin{matrix}
\boldsymbol{\varepsilon}_{k+\Delta k}^{(i),\textrm{s}} = w_N^{-\Delta k \cdot N_{\textrm{GI}}} \boldsymbol{\varepsilon}_k^{(i),\textrm{s}},\quad\quad \boldsymbol{\varepsilon}_{k+\Delta k}^{(i),\textrm{e}} = \boldsymbol{\varepsilon}_{k}^{(i),\textrm{e}},
\end{matrix} }
\label{eq:localOptHarm:transformacionesDesplazH}
\end{equation}
and the ones of the pulse in carrier $k$ that minimize the \ac{OOBE} in $\mathcal{B}_\textrm{n}^-(k+i)$ as
\begin{equation} 
\makeatletter \if@twocolumn
\boxed{ \begin{aligned}
&\boldsymbol{\varepsilon}_k^{(-i),\textrm{s}} = w_N^{-\lceil\frac{2k-N}{R}\rfloor RN_{\textrm{GI}}} \left( \boldsymbol{\varepsilon}_{k+(\lceil\frac{2k-N}{R}\rfloor R+N-2k)}^{(i),\textrm{s}} \right)^*,\\
&\boldsymbol{\varepsilon}_k^{(-i),\textrm{e}} = \left( \boldsymbol{\varepsilon}_{k+(\lceil\frac{2k-N}{R}\rfloor R+N-2k)}^{(i),\textrm{e}} \right)^*.
\end{aligned} }
\else
\boxed{ \begin{aligned}
&\boldsymbol{\varepsilon}_k^{(-i),\textrm{s}} = w_N^{-\lceil\frac{2k-N}{R}\rfloor RN_{\textrm{GI}}} \left( \boldsymbol{\varepsilon}_{k+(\lceil\frac{2k-N}{R}\rfloor R+N-2k)}^{(i),\textrm{s}} \right)^*,\quad
&\boldsymbol{\varepsilon}_k^{(-i),\textrm{e}} = \left( \boldsymbol{\varepsilon}_{k+(\lceil\frac{2k-N}{R}\rfloor R+N-2k)}^{(i),\textrm{e}} \right)^*.
\end{aligned} }
\fi \makeatother
\label{eq:localOptHarm:transformacionesSimetriaH}
\end{equation}

The matrices $\boldsymbol{\Gamma}_l^+$ and $\boldsymbol{\Gamma}_l^-$ given in (\ref{eq:localOptHarm:definicionMatrizGamma}) and (\ref{eq:localOptHarm:definicionMatrizGammaDcha}), respectively, are now obtained by replacing $\boldsymbol{\gamma}_k^{(i)}$ with $\boldsymbol{\gamma}_k^{(i),\textrm{h}}$ defined in (\ref{eq:Pgen:argminEnergiaHarm}). 

The frequency of the terms in $\mathbf{Q}_{k-i}^{+}$ depends on the spectral location of the passband edge, as Fig. \ref{fig:QmatDefinition} showed. Since these terms are located every $R$ carriers, there are $R$ different arrangements that repeat periodically. This is illustrated in Fig. \ref{fig:IDFTterms}, where edge locations that have an analogous relative position with respect to these frequency terms are depicted with the same color. These have the same residue modulo $R$, i.e., they belong to the same residue class. In that case, a relation can be established between their matrices of coefficients. For instance, $\boldsymbol{\Gamma}_{l_1}^+$ can be obtained from $\boldsymbol{\Gamma}_{l_0}^+$ by using (\ref{eq:localOptHarm:transformacionesDesplazH}) for all values of $l_1$ such that $l_1 \equiv l_0 (\textrm{mod } R)$. Fig. \ref{fig:IDFTterms} also shows that there are analogous positions between left edges (above) and right edges (below). As the orientation (left/right) of the notched band with respect to the passband edge also determines its position with respect to the frequency terms in $\mathbf{Q}_{l}^{+}$ and $\mathbf{Q}_{l}^{-}$, the color pattern depicted below is a reversed version of the one above. Hence, $\boldsymbol{\Gamma}_{l_2}^-$ can be obtained from $\boldsymbol{\Gamma}_{l_0}^+$ as long as  $l_2 \equiv -l_0 (\textrm{mod }R)$, which agrees with the relation in (\ref{eq:localOptHarm:transformacionesSimetriaH}), since $(N-l_0) \equiv -l_0 (\mathrm{mod}\;R)$.

\makeatletter \if@twocolumn
\begin{figure}[!b]
\centering
\includegraphics[width=\columnwidth]{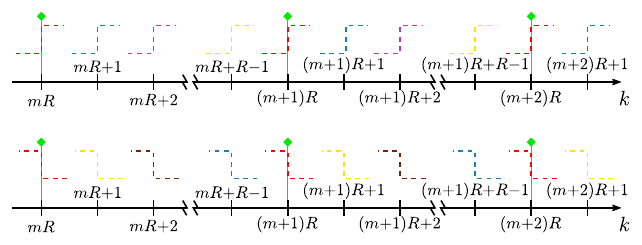}
\caption{Illustration of passband edge positions with equal relative position with respect to the \ac{IDFT} terms (in green) for left (above) and right (below) passband edges. Edge locations marked with the same colors are analogously positioned (belong to the same residue class modulo $R$) and are related by the transformations in (\ref{eq:localOptHarm:transformacionesDesplazH}) and (\ref{eq:localOptHarm:transformacionesSimetriaH}).}
\label{fig:IDFTterms}
\end{figure}
\else
\begin{figure}[!h]
\vspace{-0.4cm}
\centering
\includegraphics[width=12cm]{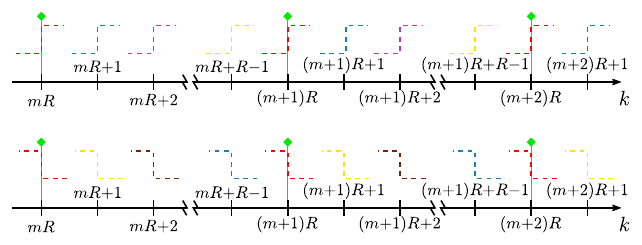}
\vspace{-0.4cm}
\caption{Illustration of passband edge positions with equal relative position with respect to the \ac{IDFT} terms (in green) for left (above) and right (below) passband edges. Edge locations marked with the same colors are analogously positioned (belong to the same residue class modulo $R$) and are related by the transformations in (\ref{eq:localOptHarm:transformacionesDesplazH}) and (\ref{eq:localOptHarm:transformacionesSimetriaH}).}
\vspace{-0.75cm}
\label{fig:IDFTterms}
\end{figure}
\fi \makeatother
\section{Spectral Shaping for Transmission Bands with Arbitrary Bandwidth: the Bandwidth-adaptive Method}
\label{Generalized_Spectral_Shaping}
A consequence of the previous transformations is that a single matrix of optimized coefficients, $\boldsymbol{\Gamma}_l^+$, suffices to obtain the matrix of coefficients required to reduce the \ac{OOBE} caused by any passband, provided that it satisfies the premise of being wide enough. However, this assumption is not suitable for most real scenarios and there are systems with narrow passbands in which the pulses have significant \ac{OOBE} over both left and right notched bands, causing their respective spectral shaping problems to be no longer independent. 
\subsection{Pulse Definition and Optimization}
To shape signals with narrow passbands, additional \ac{AIC} and \ac{AST} terms must be added to the pulse  proposed in (\ref{eq:Pgen:pGenFlIzq}) and the distance to both edges of the passband must be stated, yielding
\begin{equation}
\mathbf{h}_k^{(i,j)} = \mathbf{p}_k + \mathbf{C}_{k-i}^+ \boldsymbol{\alpha}_k^{(i)} + \mathbf{t}_k^{(i)} + \mathbf{C}_{k-j}^- \boldsymbol{\alpha}_k^{(j)} + \mathbf{t}_k^{(j)}, \quad i>0, j<0 
\label{eq:BWadapt:pulsGenNuevo}
\end{equation}
where $i>0$ denotes that carrier $k$ is the $i$-th one above the left edge of the passband and $j<0$ that it is the $j$-th one below the right edge. The coefficients in (\ref{eq:BWadapt:pulsGenNuevo}) have to be jointly optimized to minimize the \ac{OOBE} at the notched bands located by the passband left and right edges.

%
Since the width of the passband determines the strength of the influence between the shaping problems located at its edges, it becomes a new parameter to be considered in the optimization procedure. Let $N_{\textrm{D}}$ be the number of carriers of the passband (those located right on the edges are not considered in-band). The optimization procedure presented in this work considers a set of passbands ranging from $N_{\textrm{D\_min}} = N_{\textrm{h\_min}}+2N_{\textrm{ci}}$ to $N_{\textrm{D\_max}} = N_{\textrm{h\_max}}+2N_{\textrm{ci}}$, where $N_{\textrm{D\_min}}$ is the narrowest passband in which spectral shaping would be performed and from which $N_{\textrm{h\_min}}$ is determined. The value of $N_{\textrm{h\_max}}$ is selected such that $N_{\textrm{h}} > N_{\textrm{h\_max}}$ yields a negligible \ac{OOBE} reduction. Fig. \ref{fig:passbandWidths} depicts the set of considered passbands, where the \ac{CC} are color-coded in yellow, and the data carriers (DC) are in blue. Notice that $N_{\textrm{D}} = N_{\textrm{h}} + 2N_{\textrm{ci}}$ in all cases, meaning that all data carriers in the passband make use of the pulse proposed in (\ref{eq:BWadapt:pulsGenNuevo}) to minimize the \ac{OOBE} in both notched bands at the same time. To emphasize this end, the indices $(i, j)$ are displayed above each data carrier.

\makeatletter \if@twocolumn
\begin{figure}[!t]
\centering
\includegraphics[width=\columnwidth]{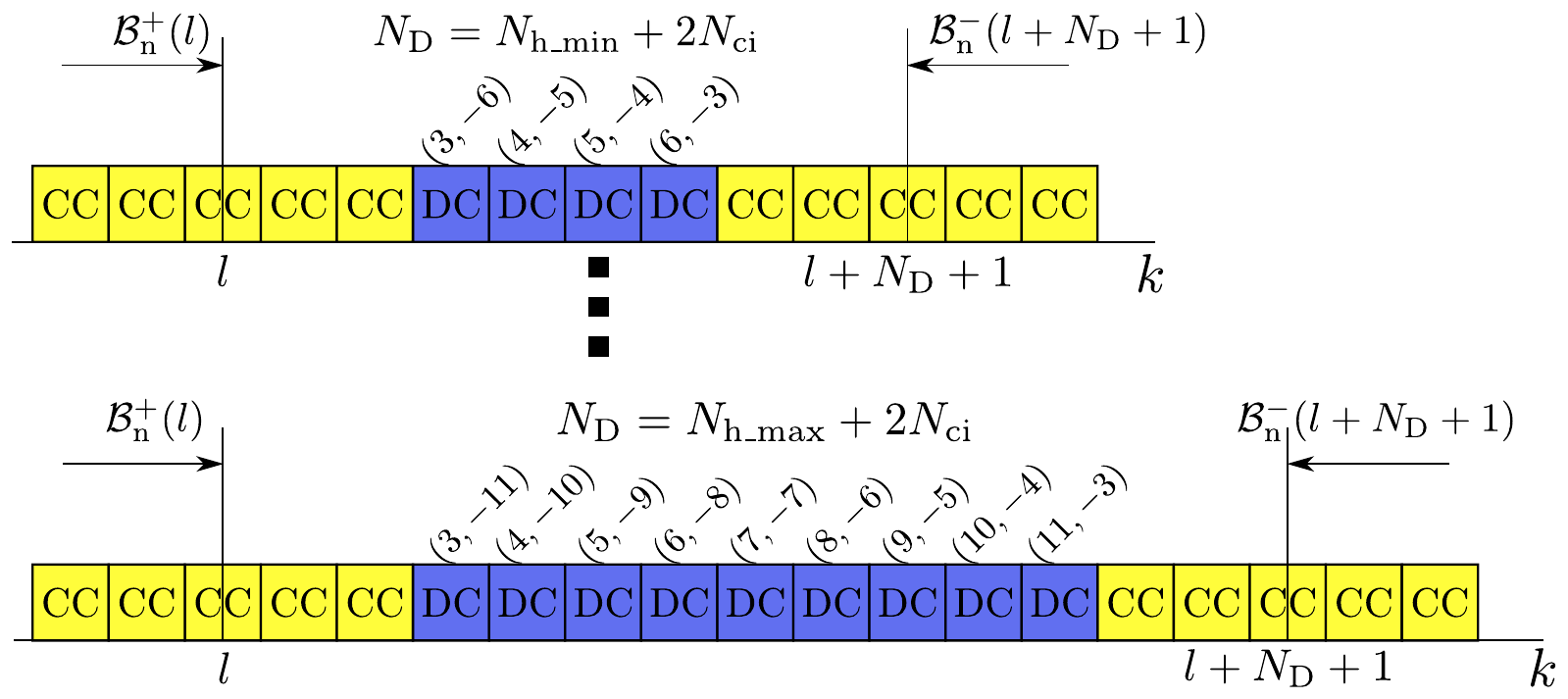}
\caption{Passbands of widths ranging from $N_{\textrm{D}\_\textrm{min}}$ to $N_{\textrm{D}\_\textrm{max}}$.}
\label{fig:passbandWidths}
\end{figure}
\else
\begin{figure}[!h]
\centering
\includegraphics[width=10.5cm]{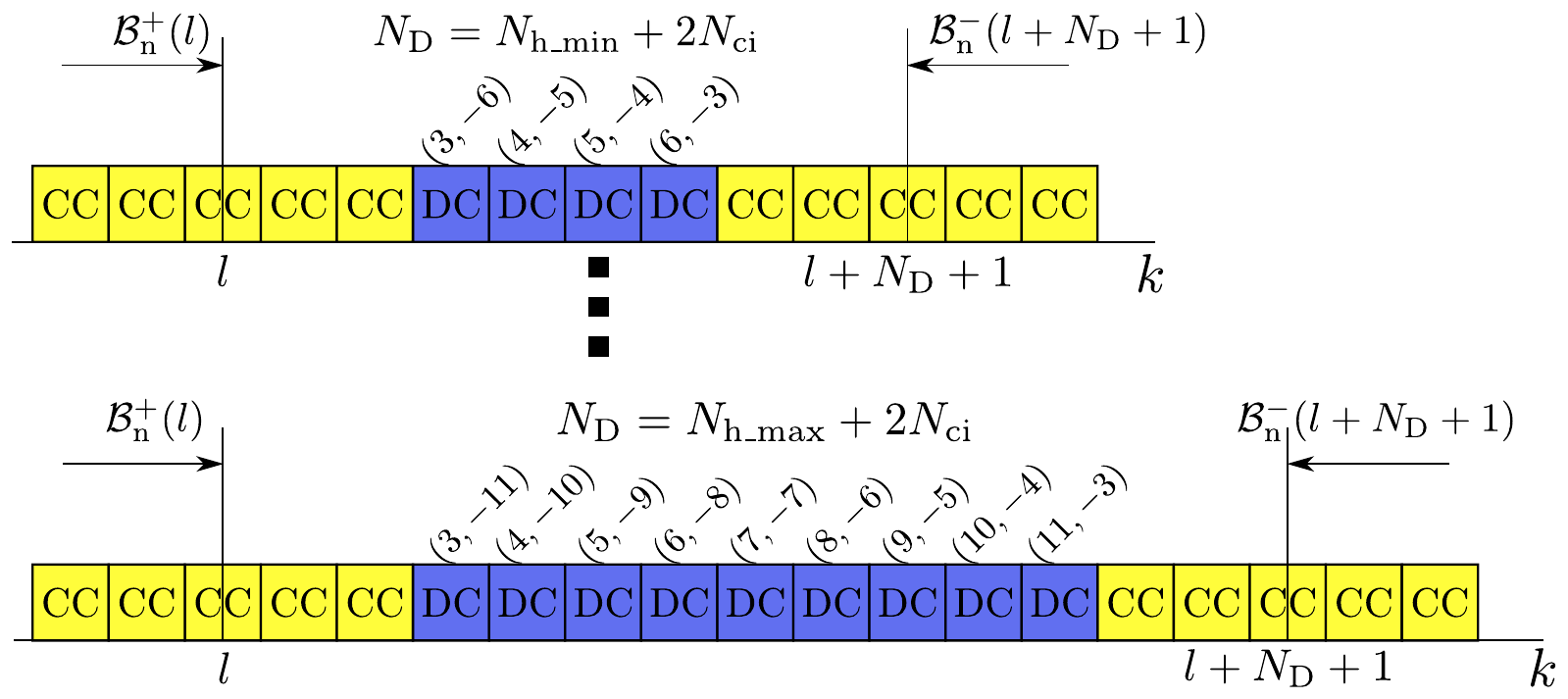}
\vspace{-0.4cm}
\caption{Passbands of widths ranging from $N_{\textrm{D}\_\textrm{min}}$ to $N_{\textrm{D}\_\textrm{max}}$.}
\vspace{-0.75cm}
\label{fig:passbandWidths}
\end{figure}
\fi \makeatother

The energy emitted by $\mathbf{h}_k^{(i,j)}$ to the left and right notched bands can be expressed as
\begin{equation} 
\begin{aligned}
E_k^{(i,j)} = (\mathbf{h}_k^{(i,j)})^{\textrm{H}} \left(\boldsymbol{\Phi}_{\mathcal{B}_{\textrm{n}}^+(k-i)}  +  \boldsymbol{\Phi}_{\mathcal{B}_{\textrm{n}}^-(k-j)}\right) \mathbf{h}_k^{(i,j)}.
\end{aligned}
\label{eq:BWadapt:energiaOOBE}
\end{equation}

The following cost function associated to the passband with left edge at $l$ and width $N_{\textrm{D}}$
\begin{equation}
F_l^{(N_{\textrm{D}})} = \sum_{m=1}^{N_{\textrm{h}}} E_{l+N_{\textrm{ci}}+m}^{(N_{\textrm{ci}}+m,-N_{\textrm{D}}-1+N_{\textrm{ci}}+m)},
\label{eq:BWadapt:funCosteBandakNd}
\end{equation}
is the summation of the energy emitted to the notched bands by each of the proposed pulses. It depends on the coefficients in the matrices $\boldsymbol{\Gamma}_l^+=\left[ \boldsymbol{\gamma}_{l+N_{\textrm{ci}}+1}^{(N_{\textrm{ci}}+1)}, \cdots, \boldsymbol{\gamma}_{l+N_{\textrm{ci}}+N_{\textrm{h\_max}}}^{(N_{\textrm{ci}}+N_{\textrm{h\_max}})} \right]$ and $\boldsymbol{\Gamma}_{l+N_\textrm{D}+1}^-=\left[\boldsymbol{\gamma}_{l+N_{\textrm{ci}}+N_{\textrm{h\_max}}}^{(-N_{\textrm{D}}-1+N_{\textrm{ci}}+N_{\textrm{h\_max}})}, \right.$ $\left.\cdots, \boldsymbol{\gamma}_{l+N_{\textrm{ci}}+1}^{(-N_{\textrm{D}}-1+N_{\textrm{ci}}+1)} \right]$. However, since the coefficients in the first and the second matrices are related through the transformations (\ref{eq:localOpt:transformacionesDesplazamiento}) and (\ref{eq:localOpt:transformacionesGiro}),  $\boldsymbol{\gamma}_k^{(-i)} = f(\boldsymbol{\gamma}_k^{(i)})$, the cost function depends only on the first matrix of coefficients, $F_l^{(N_{\textrm{D}})}\left(\boldsymbol{\Gamma}_l^+\right)$.

The aggregate \ac{OOBE} in all the considered passband widths is then given by
\begin{equation}
F_l \left( \boldsymbol{\Gamma}_l^+ \right) = \sum_{n_\textrm{D} = N_{\textrm{D\_min}}}^{N_{\textrm{D\_max}}} F_l^{(n_\textrm{D})} \left( \boldsymbol{\Gamma}_l^+ \right),
\label{eq:BWadapt:funCosteSumaBandas}
\end{equation}
and the optimal coefficients are obtained as
\begin{equation}
\widehat{\boldsymbol{\Gamma}}_l^+ = \argmin_{\boldsymbol{\Gamma}_l^+}\left\{ F_l \left( \boldsymbol{\Gamma}_l^+ \right) \right\}.
\label{eq:BWadapt:argminFunCoste}
\end{equation}

\subsection{Transition Pulses with Harmonically Designed Boundaries}
\label{Transition_Pulses_with_Harmonically_Designed_Boundaries}
When the transition pulses are harmonically designed, the complexity of the proposed optimization strategy escalates, as $R$ different matrices of coefficients like the one in (\ref{eq:localOptHarm:definicionMatrizGamma}) have to be obtained, one for each relative position of the passband edge with respect to the \ac{IDFT} terms (see Fig. \ref{fig:IDFTterms}). Each of these matrices should belong to a different residue class modulo $R$. To that end, they can have consecutive subindices, ranging from $l$ to $l+R-1$. Let
\begin{equation}
\boldsymbol{\Upsilon}_l^+ = \left\{ \boldsymbol{\Gamma}_l^+, \boldsymbol{\Gamma}_{l+1}^+, \dots, \boldsymbol{\Gamma}_{l+R-1}^+ \right\}
\label{eq:BWadapt:defConjuntoY}
\end{equation}
be the set containing the matrices of complex coefficients for all residue classes modulo $R$.

The coefficients associated to the right edge of any of the considered passbands can be expressed as a function of one of those in the set (\ref{eq:BWadapt:defConjuntoY}). Let $F_{l+r}^{(N_{\textrm{D}})} \left( \boldsymbol{\Gamma}_{l+r}^+, \boldsymbol{\Gamma}_{l+r+N_{\textrm{D}}+1}^- \right)$ be the cost function associated to any shifted version of the passbands in Fig. \ref{fig:passbandWidths}, whose left and right edges are located at $l+r$ and $l+r+N_{\textrm{D}}+1$, respectively. The matrix $\boldsymbol{\Gamma}_{l+r+N_{\textrm{D}}+1}^-$ can be expressed as a function of $\boldsymbol{\Gamma}_{N-(l+r+N_{\textrm{D}}+1)}^+$ and, likewise, the latter can be expressed as a function of another matrix of coefficients, $\boldsymbol{\Gamma}_{l+r'}^+$. Consequently, the cost function dependency can be rewritten as $F_{l+r}^{(N_{\textrm{D}})} \left( \boldsymbol{\Gamma}_{l+r}^+, \boldsymbol{\Gamma}_{l+r+N_{\textrm{D}}+1}^- \right) = F_{l+r}^{(N_{\textrm{D}})} \left( \boldsymbol{\Gamma}_{l+r}^+, \boldsymbol{\Gamma}_{l+r'}^+ \right)$, where $r' = [N-(2l+r+N_{\textrm{D}}+1)] \textrm{ mod } R$ leads to $\boldsymbol{\Gamma}_{l+r'}^+$ being contained in the set $\boldsymbol{\Upsilon}_l^+$. Therefore, the cost functions associated to each of the shifted versions of the passbands in Fig. \ref{fig:passbandWidths} ultimately depend on the matrices of coefficients contained in $\boldsymbol{\Upsilon}_l^+$. Finally, the set of optimal coefficients is computed as
\begin{equation}
\widehat{\boldsymbol{\Upsilon}}_l^{+} = \argmin_{\boldsymbol{\Upsilon}_l^+} \left\{ \sum_{r=0}^{R-1} F_{l+r}\left( \boldsymbol{\Gamma}_{l+r}^+ \right) \right\}.
\label{eq:BWadapt:argminUpsilon}
\end{equation}

Both (\ref{eq:BWadapt:argminFunCoste}) and (\ref{eq:BWadapt:argminUpsilon}) should include constraints for the absolute value of the real and imaginary parts of the coefficients involved to avoid \ac{PSD} peaks in the passband.

\subsection{Computational Cost and Memory Requirements of the Proposed Method}
\label{Computational cost}
The proposed method consists of three phases: first, the original matrix (or matrices) of optimal coefficients are obtained offline through an optimization procedure. Hence, its computational cost is irrelevant for the communication device. Second, these matrices of coefficients are stored in the communication device and transformed online to adapt them to the actual passband edges of the transmission band. Two different schemes can be adopted in this phase: 

\renewcommand{\theenumi}{\Alph{enumi})}%
\begin{enumerate}
\item Compute the transformed coefficients just once at the beginning and every time the transmission mask changes. The resulting sets of coefficients are stored in the transmitter until a change in the \ac{PSD} mask occurs.
\item Compute the transformed coefficients for every transmitted symbol, saving memory at the cost of higher computational cost. 
\end{enumerate}

Table \ref{Table:CosteCompExpresiones} contains the expressions of the computational cost and memory requirements (per \ac{OFDM} symbol) associated to the second phase for both proposed schemes. The former is expressed in number of complex products and the latter in number of stored complex coefficients. Only the proposed pulse in (\ref{eq:BWadapt:pulsGenNuevo}) is considered, as it is the worst-case scenario in terms of memory and complexity. Similarly, only harmonically designed transition pulses are assessed, since they attain similar \ac{OOBE} reduction to the regular ones but have lower complexity \cite{Diez19}.

In scheme A, the computational cost associated to the transformations applied to adapt the precomputed proposed pulses to the ones employed in the considered scenario has been neglected because they are performed only when the transmission mask changes, and it is reasonable to assume that the frequency of these changes is far below the frequency at which \ac{OFDM} symbols are transmitted. Scheme B does not require storing that many coefficients, although there is some additional computational cost in the transmission of every \ac{OFDM} symbol that corresponds to the transformation of the harmonically designed transition pulses.

\makeatletter \if@twocolumn
\begin{table}[!ht]
\centering
\renewcommand{\arraystretch}{1.4}
\caption{Computational cost and memory requirements per \ac{OFDM} symbol}
\vspace{-0.4cm}
\begin{tabular}{C{1.4cm} C{3cm} C{3.3cm}} 
\cline{2-3}
         & Number of complex products & Number of complex coefficients to be stored\\ \hline
 Scheme A & $-$ & $2|\mathcal{D}^{\textrm{h}}|(N_{\textrm{CC}} + 2N_{\textrm{QQ}})$  \\
 Scheme B & $2|\mathcal{D}^{\textrm{h}}| N_{\textrm{QQ}}$ & $N_{\textrm{h}\_\textrm{max}}R(N_{\textrm{CC}} + 2N_{\textrm{QQ}})$ \\
\hline
\end{tabular}
\label{Table:CosteCompExpresiones}
\vspace{-0.4cm}
\end{table}
\else
\begin{table}[!ht]
\centering
\renewcommand{\arraystretch}{1.3}
\caption{Computational cost and memory requirements per \ac{OFDM} symbol}
\begin{tabular}{C{1.4cm} C{5cm} C{5.8cm}} 
\cline{2-3}
         & Number of complex products & Number of complex coefficients to be stored\\ \hline
 Scheme A & $-$ & $2|\mathcal{D}^{\textrm{h}}|(N_{\textrm{CC}} + 2N_{\textrm{QQ}})$  \\
 Scheme B & $2|\mathcal{D}^{\textrm{h}}| N_{\textrm{QQ}}$ & $N_{\textrm{h}\_\textrm{max}}R(N_{\textrm{CC}} + 2N_{\textrm{QQ}})$ \\
\hline
\end{tabular}
\label{Table:CosteCompExpresiones}
\vspace{-0.4cm}
\end{table}
\fi \makeatother

Finally, in the third phase, the samples of the \ac{OFDM} symbol are computed using the transformed coefficients. The computational cost of this process is omitted, as it can be found in \cite{Diez19}. The total cost of the proposed schemes is obtained by adding the latter to the values in Table \ref{Table:CosteCompExpresiones}. 


The implementation complexity (cost of the second and third phases) depends on the number of carriers that employ the proposed pulse, which can be reduced at the expense of degrading the performance. While the cost of these phases is lower in other methods that use the same pulse in all carriers, e.g., \cite{Mahmoud08}, \cite{Brandes10} and {\color{reviewer1} \cite{Hussain2022}}, the key difference is that they have to perform the first phase for each \ac{OFDM} symbol or, at best, whenever the emission mask changes. In contrast, in the proposed method it is performed offline and only once, which yields a lower overall cost, since solving the optimization problem is much more costly than the operations in Table \ref{Table:CosteCompExpresiones}. 


\section{Numerical Results}
\label{Numerical_results}
For illustrative purposes, an \ac{OFDM} system like the one defined in the ITU-T Rec. G.9960 is employed \cite{G9960}. It uses $N=4096$, $N_{\textrm{GI}}=1024$ and $\beta=512$. The $\beta$ samples at both ends of $g(n)$ are shaped using a \ac{RC} window. The proposed pulses that only use \ac{CC} are denoted as $h_k(\textrm{CC})$, while the ones with \ac{CC} and harmonically designed transition pulses are designated as $h_k(\textrm{CC}, t_k\textrm{-h})$. The one with \ac{CC} and regular transition pulses is not considered in this section due to its high computational cost. {\color{reviewer1} The real and imaginary parts of the coefficients have been constrained to $\sqrt{2}$ to avoid \ac{PSD} peaks in the passband.}

Three different pulse design methods are compared. The one in \cite{Diez19}, which yields the highest performance as pulses are designed \textit{ad hoc} for the considered scenario. The \textit{local optimization} method presented in Section \ref{Problem_statement}, in which a large passband bandwidth is assumed and the set of precomputed pulses have been locally optimized to minimize the \ac{OOBE} only in the notched band closest to them, and the \textit{bandwidth adaptive} method proposed in Section \ref{Generalized_Spectral_Shaping}, in which the precomputed pulses have been optimized to minimize the aggregate \ac{OOBE} of a set of narrow passbands. The figure of merit used for comparison is the largest value of the normalized \ac{PSD} in the notched band, denoted as $\textrm{PSD}_{\textrm{max}}$, which is typically attained in the closest frequencies to the passband edges. The \ac{PSD} is analytically computed using (\ref{eq:bgnd:PSDsenalOFDM}) in all cases. 

\subsection{Influence of the Passband Width in the Performance}

In this section the performance of the proposed methods is evaluated in a wide passband of $N_{\textrm{D}}=45$ carriers, where the left and right notched bands comprise the carrier indexes given by $\mathcal{B}_{\textrm{n}}^+(100)=\{k| \; 0 \leq k <100\}$ and $\mathcal{B}_{\textrm{n}}^-(146)=\{k| \;  146< k \leq 4095\}$, respectively, and in a narrow passband of $N_{\textrm{D}}=8$ carriers, with $\mathcal{B}_{\textrm{n}}^+(100)=\{k|\; 0\leq k <100\}$ and $\mathcal{B}_{\textrm{n}}^-(109)=\{k|\; 109 < k \leq 4095\}$, respectively. The width $N_{\textrm{D}}=45$ has been empirically determined as the minimum number of data carriers required for the proposed pulses employed at one edge of the passband to not interfere with those used at the opposite edge.


Fig. \ref{fig:PSDnarrowAICAIChtk} depicts the normalized \ac{PSD} obtained in both cases when the optimized pulses $h_k(\textrm{CC}, t_k\textrm{-h})$ are determined using the \textit{ad hoc} \cite{Diez19}, \textit{local optimization} and \textit{bandwidth adaptive} methods. As  reference, the normalized \ac{PSD} obtained when only \ac{RC} pulse-shaping is employed is also depicted {\color{reviewer1}(note that curves for $N_{\textrm{D}}=45$ and $N_{\textrm{D}}=8$ fully overlap)}. The optimized pulses are configured with $N_{\textrm{ci}}=2$, $N_{\textrm{co}}=2$ and $N_{\textrm{q}}=2$ in the three methods. The number of data carriers with proposed pulses is $N_{\textrm{h}}=13$ in the wide passband and $N_{\textrm{h}}=4$ in the narrow one (i.e., all data carriers use optimized pulses, as $N_{\textrm{D}}=N_{\textrm{h}}+2N_{\textrm{ci}}$).

The pulses of the \textit{local optimization} method have been obtained for a passband width of $N_{\textrm{D}}=45$ carriers (the minimum for it to be considered wide) whose left notched band is $\mathcal{B}_{\textrm{n}}^+(4050) = \{ k| \; 0 \leq k < 4050 \}$, i.e. the right edge is located at the right end of the spectrum. Only the pulses employed to shape the left edge have been determined. These are then transformed by means of (\ref{eq:localOpt:transformacionesDesplazamiento}) and (\ref{eq:localOpt:transformacionesGiro}) and applied to the left edge of the scenario in Fig. \ref{fig:PSDnarrowAICAIChtk} and by means of (\ref{eq:localOptHarm:transformacionesDesplazH}) and (\ref{eq:localOptHarm:transformacionesSimetriaH}) and applied to the right edge of this scenario (not shown in the figure). The pulses of the \textit{bandwidth adaptive} technique have been determined  in a scenario where the left passband edge is located at $f=1/2$ and the right edge is shifted to the right as the passband is enlarged from $N_{\textrm{h\_min}}=4$ to $N_{\textrm{h\_max}}=13$ (see Fig. \ref{fig:passbandWidths}). The $R=N/\beta=8$ sets of transition pulses with harmonically designed boundaries have been obtained by displacing the left edge of the passband from $f=\left\{\frac{1}{2},\frac{1}{2}+\frac{1}{N}, \ldots, \frac{1}{2}+\frac{R-1}{N}\right\}$. The resulting optimized pulses are then transformed and applied to the scenario in Fig. \ref{fig:PSDnarrowAICAIChtk} by using (\ref{eq:localOpt:transformacionesDesplazamiento}),  (\ref{eq:localOpt:transformacionesGiro}), (\ref{eq:localOptHarm:transformacionesDesplazH}) and (\ref{eq:localOptHarm:transformacionesSimetriaH}).

As seen, the \textit{local optimization} method yields $\textrm{PSD}_{\textrm{max}}=-55.3$ dB when $N_{\textrm{D}}=45$, which is almost the same as the value given by the \textit{ad hoc} one. This indicates that this moderate passband width, which comprises just $1.1\%$ of the system carriers, is wide enough for both the closest carriers to the passband edge that employ conventional pulses and the proposed pulses closest to the opposite edge to have negligible \ac{OOBE} in $\mathcal{B}_{\textrm{n}}^+(100)$. The gain of these methods with respect to \ac{RC} pulse-shaping exceeds $42$ dB. The \textit{bandwidth adaptive} method gives $\textrm{PSD}_{\textrm{max}}=-48.2$ dB. While this is a notable reduction with respect to the \ac{RC} pulse-shaping, it is about $7$ dB worse than the value attained by the \textit{local optimization}. However, when $N_{\textrm{D}}=8$, the latter performs very poorly, while the \textit{bandwidth adaptive} method gives $\textrm{PSD}_{\textrm{max}}=-45.3$ dB, which is only $2$ dB larger than the one attained by the \textit{ad hoc} solution. As a reference, the \ac{ACLR} limit set in \ac{5G} is $45$ dB \cite{TS38.104}. Hence, the \textit{bandwidth adaptive} method provides a notable \ac{OOBE} reduction in all passband widths. 


\makeatletter \if@twocolumn
\begin{figure}[!t]
\centering
\includegraphics[width=\columnwidth]{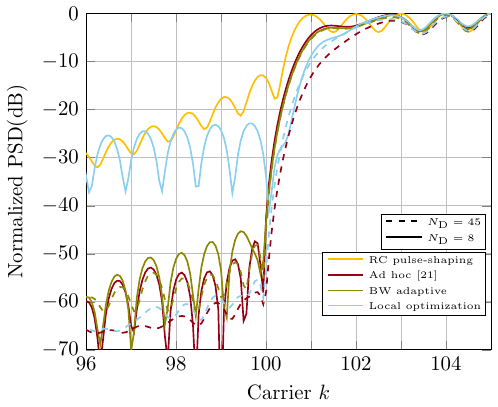}
\caption{Normalized \ac{PSD} in a wide passband with $N_{\textrm{D}}=45$ and a narrow one with $N_{\textrm{D}}=8$. Spectral shaping is performed by means of $h_k(\mathrm{CC}, t_k\textrm{-}h)$ designed using the \textit{ad hoc} \cite{Diez19},  \textit{local optimization} and \textit{bandwidth adaptive} methods. The \ac{OFDM} system with \ac{RC} window is shown as a reference.}
\label{fig:PSDnarrowAICAIChtk}
\vspace{-0.5cm}
\end{figure}
\else
\begin{figure}[!h]
\centering
\includegraphics[width=8.5cm]{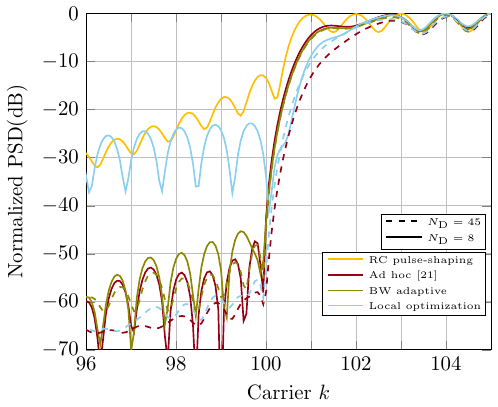}
\vspace{-0.5cm}
\caption{Normalized \ac{PSD} in a wide passband with $N_{\textrm{D}}=45$ and a narrow one with $N_{\textrm{D}}=8$. Spectral shaping is performed by means of $h_k(\mathrm{CC}, t_k\textrm{-}h)$ designed using the \textit{ad hoc} \cite{Diez19},  \textit{local optimization} and \textit{bandwidth adaptive} methods. The \ac{OFDM} system with \ac{RC} window is shown as a reference {\color{reviewer1}(curves for $N_{\textrm{D}}=45$ and $N_{\textrm{D}}=8$ fully overlap).}}
\label{fig:PSDnarrowAICAIChtk}
\vspace{-0.4cm}
\end{figure}
\fi \makeatother
As expected, reducing the \ac{OOBE} is more challenging in narrow passbands, since the contribution of each carrier to the \ac{OOBE} in both notched bands is greater. However, the particularly poor performance of the \textit{local optimization} is due to the emission of the proposed pulses located closest to the right edge of the passband, which have been optimized to minimize the \ac{OOBE} for a wide passband of at least $N_{\textrm{D}}=45$. Since this method has a more limited validity than the \textit{bandwidth adaptive} one, it will not be considered in the next sections. 
It has been verified (although not shown in Fig. \ref{fig:PSDnarrowAICAIChtk}) that the additional \ac{OOBE} reduction given by the outband \ac{CC} in the \textit{bandwidth adaptive} method is very small. Hence, $N_{\textrm{co}}=0$ will be used hereafter in order to reduce the computational cost. The performance of the \textit{bandwidth adaptive} technique in wide passbands can be improved by setting a larger $N_{\textrm{h\_min}}$, at the cost of deteriorating the performance in narrow passbands. A detailed analysis is presented in the following subsection. 

\subsection{Influence of the Method Parameters in the Performance}
The performance of the \textit{bandwidth adaptive} method in narrow passbands depends on the parameters $N_{\textrm{h\_min}}$ and $N_{\textrm{h\_max}}$ employed in the optimization procedure. This section assesses their influence in passbands as narrow as $N_{\textrm{D}} = 5$. Representative examples of this worst-case scenario are the \ac{NB-IoT} feature of \ac{LTE} and of some transmission subbands between consecutive notches defined in the EN 50561-1 \cite{Ratasuk2017}\cite{EN50561-1}. In the forthcoming results, $N_{\textrm{ci}}=2$, $N_{\textrm{co}}=0$ and $N_{\textrm{q}} = 2$ have been employed.

\begin{figure}[!t]
\centering
\vspace{-0.2cm}
\makeatletter \if@twocolumn
\includegraphics[width=\columnwidth]{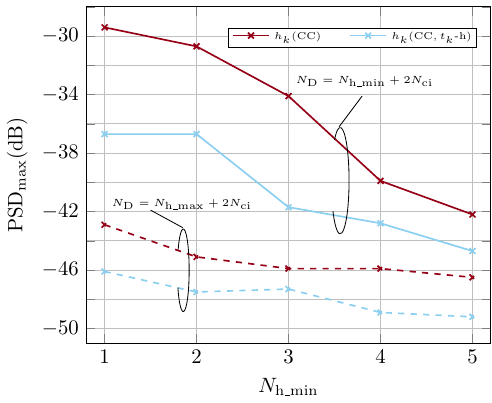}
\else
\includegraphics[width=8.5cm]{./Fig/PSD_vs_Nhmin.pdf}
\fi \makeatother
\vspace{-0.7cm}
\caption{$\mathrm{PSD}_{\mathrm{max}}$ attained by $h_k(\textrm{CC})$ and $h_k(\textrm{CC}, t_k\textrm{-h})$ when determined with the \textit{bandwidth adaptive} method in a passband width of $N_{\textrm{D}}=N_{\textrm{h}\_\textrm{max}}+2N_{\textrm{ci}}$ carriers and in the worst case $N_{\textrm{D}}=N_{\textrm{h}\_\textrm{min}}+2N_{\textrm{ci}}$.}
\label{fig:PSDvsNhmin}
\vspace{-0.8cm}
\end{figure}

Fig. \ref{fig:PSDvsNhmin} shows the $\textrm{PSD}_{\textrm{max}}$ attained in the notched band for $N_{\textrm{h\_min}}=\{1,2,\ldots,5\}$ and $N_{\textrm{h\_max}}=13$. Optimized pulses with two types of \ac{OOBE} reduction terms are considered: $h_k(\textrm{CC})$, depicted in red, and $h_k(\textrm{CC},t_k\textrm{-h})$, drawn in light blue. These are applied to two different types of passband: the narrowest passband for each $N_{\textrm{h\_min}}$, i.e., $N_{\textrm{D}}=N_{\textrm{h\_min}}+2N_{\textrm{ci}}$, in solid line, and a wider passband of $N_{\textrm{D}} = N_{\textrm{h\_max}}+2N_{\textrm{ci}}$ carriers, in dashed line. 

Fig. \ref{fig:PSDvsNhmin} shows that increasing the value of $N_{\textrm{h\_min}}$ leads to a better \ac{OOBE} reduction both for $h_k(\textrm{CC})$ and $h_k(\textrm{CC},t_k\textrm{-h})$. The selection of $N_{\textrm{h\_min}}$ involves a trade-off between the minimum \ac{OOBE} attained in narrow passbands and that obtained in wide passbands. Increasing $N_{\textrm{h\_min}}$ reduces the range of bandwidths to be considered in the optimization in (\ref{eq:BWadapt:argminFunCoste}), which relaxes the referred trade-off, and leads to better performance in the passbands considered in the optimization. However, such solutions yield a suboptimal \ac{OOBE} reduction when $N_{\textrm{D\_min}} \leq N_{\textrm{h\_min}}+2N_{\textrm{ci}}$.

As expected, $h_k(\textrm{CC},t_k\textrm{-h})$ outperforms $h_k(\textrm{CC})$, since the former has more degrees of freedom. It can be seen that $N_{\textrm{D}}=N_{\textrm{h\_max}}+2N_{\textrm{ci}}$ yields lower $\textrm{PSD}_{\textrm{max}}$ than $N_{\textrm{D}}=N_{\textrm{h\_min}}+2N_{\textrm{ci}}$, but the rate at which it decreases as $N_{\textrm{h\_min}}$ increases is more pronounced in the latter than in the former. This is because the wider the passband, the lower the \ac{OOBE} emitted by the optimized pulses at the farthest edge of the passband. Hence, widening a large passband yields little \ac{OOBE} reduction. 

The value of $N_{\textrm{h\_min}}$ is determined to ensure that the \ac{OOBE} level attained in the narrowest passband is below the required limit. Fig. \ref{fig:PSDvsNhmin} can be used to perform this selection. For instance, in order to guarantee that $\textrm{PSD}_{\textrm{max}}\leq -41$ dB  with $h_k(\textrm{CC})$, $N_{\textrm{h\_min}} \geq 5$ must be employed, while with $h_k(\textrm{CC}, t_k\textrm{-h})$ this can be attained with $N_{\textrm{h\_min}} \geq 3$. 

\begin{figure}[!t]
\centering
\vspace{-0.2cm}
\makeatletter \if@twocolumn
\includegraphics[width=\columnwidth]{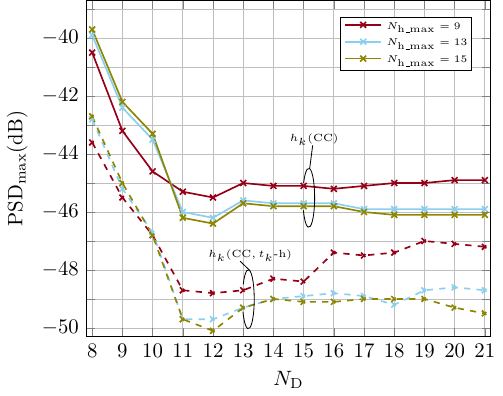}
\else
\includegraphics[width=8.5cm]{./Fig/PSD_vs_Nhmax_v2.pdf}
\fi \makeatother
\vspace{-0.7cm}
\caption{$\mathrm{PSD}_{\mathrm{max}}$ attained by $h_k(\textrm{CC})$ and $h_k(\textrm{CC}, t_k\textrm{-h})$ in passband widths $8\leq N_{\textrm{D}} \leq 21$. The proposed pulses are designed with the \textit{bandwidth adaptive} method parameterized with $N_{\textrm{h}\_\textrm{min}} = 4$ and $N_{\textrm{h}\_\textrm{max}}=\{9, 13, 15\}$.}
\label{fig:PSDvsNhmax}
\vspace{-0.8cm}
\end{figure}

In order to assess the influence of $N_{\textrm{h\_max}}$ in the performance, Fig. \ref{fig:PSDvsNhmax} depicts the values of $\textrm{PSD}_{\textrm{max}}$ attained by $h_k(\textrm{CC})$ and $h_k(\textrm{CC}, t_k\textrm{-h})$ when designed using the \textit{bandwidth adaptive} method in passbands widths ranging from $N_{\textrm{D}}=8$ to $N_{\textrm{D}} = 21$. Solutions obtained for $N_{\textrm{h\_max}}=\{9, 13, 15\}$ and $N_{\textrm{h\_min}} = 4$ are displayed. Note that the number of carriers with optimized pulses is $N_{\textrm{h}}=\max\left(N_{\textrm{D}}-2N_{\textrm{ci}},N_{\textrm{h\_max}}\right)$. Two regions can be distinguished: a decreasing one from $N_{\textrm{D}}=8$ to $N_{\textrm{D}}=11$ which is followed by an almost flat region. In the former, the $\textrm{PSD}_{\textrm{max}}$ decreases as $N_{\textrm{D}}$ increases because widening the passband reduces the emissions received in each notched band from the optimized pulses employed in the farthest passband edge. This trend continues until the lowest \ac{OOBE} level that the computed solutions can achieve is reached. 

As seen, $\textrm{PSD}_{\textrm{max}}$ values achieved with $h_k(\textrm{CC},t_k\text{-h})$ are about $3$ dB lower than with $h_k(\textrm{CC})$ in almost the whole range of $N_{\textrm{D}}$, except for $N_{\textrm{h\_max}}=9$, where differences are smaller. Moreover, in the latter, $\textrm{PSD}_{\textrm{max}}$ clearly increases for $N_{\textrm{D}}>15$ instead of stabilizing around a certain value, as other curves do. This is due to $N_{\textrm{h\_max}}=9$ being an insufficient number of data carriers employing the optimized pulse. Hence, the reduction attained by using these pulses in the $N_{\textrm{h\_max}}=9$ data carriers located closest to each edge is spoiled by the emission of the carriers situated consecutive to the $9^{\mathrm{th}}$ one, which use conventional pulses. This phenomenon does not manifest in the $h_k(\textrm{CC})$ cases because they achieve a more limited \ac{OOBE} reduction.

Fig. \ref{fig:PSDvsNhmax} can be used to determine the most suitable value for $N_{\textrm{h\_max}}$ in a given application attending to the \ac{OOBE} requirements. For instance, results attained with $N_{\textrm{h\_max}}=9$ are outperformed by the ones obtained with $N_{\textrm{h\_max}}=13$ and $N_{\textrm{h\_max}}=15$ in wide passbands, but the three of them achieve very similar $\mathrm{PSD}_{\mathrm{max}}$ values for $N_{\textrm{D}} < 10$. It can be also noticed that  $N_{\textrm{h\_max}}=13$ seems to be an appropriate value for this parameter, since it gives almost the same performance as $N_{\textrm{h\_max}}=15$ and is slightly more energy-efficient and computationally simpler.

\subsection{Comparison with Previous Methods}
{\color{reviewer1} The \textit{bandwidth adaptive} method is now compared with the time-domain method by Mahmoud et al. \cite{Mahmoud08} and the methods that combine \ac{AIC} and time-domain strategies by Brandes \textit{et al.} \cite{Brandes10}, Hussain et al. \cite{Hussain2022} and the \textit{ad hoc} one in \cite{Diez19}.} The scenario with notched bands at carrier indexes $\mathcal{B}_{\textrm{n}}=\{ 0, \dots, 1024,\allowbreak 3022,$ $\dots, 3026, 3072, \dots, 4095 \}$ given in \cite{Diez19} is considered. 


The \ac{PSD} obtained with the methods in \cite{Brandes10} and \cite{Mahmoud08} is computed by applying the Welch's averaged periodogram method with a 16384-sample Hanning window and $4096$-sample overlap to an \ac{OFDM} signal consisting of $2000$ QPSK modulated symbols. {\color{reviewer1} The \ac{PSD} for the method in \cite{Hussain2022} is obtained analytically using  \cite[eq. (10)]{Hussain2022}}. Optimized pulses in the \textit{bandwidth adaptive} and \textit{ad hoc} \cite{Diez19} methods are designed with $N_{\textrm{ci}}=2$, $N_{\textrm{co}}=0$ and $N_{\textrm{q}} = 2$. {\color{reviewer1} Likewise, the method in \cite{Hussain2022} is configured to use the same set of \ac{CC} as the aforementioned techniques and the tapered transitions also span $\beta$ samples. Its solution is obtained after 16 iterations and the regularization term is configured to avoid \ac{PSD} peaks in the passband.} For the \textit{bandwidth adaptive} method, $N_{\textrm{h\_min}}=4$ and $N_{\textrm{h\_max}}=13$ have been chosen. The \textit{ad hoc} design also uses $N_\textrm{h}=13$. While this method can give lower \ac{PSD} values by using a larger $N_\textrm{h}$, as shown in \cite[Fig. 4]{Diez19}, this will increase the computational cost of the third phase described in Section \ref{Computational cost}, since a larger number of optimized pulses would be employed.  

\begin{figure}[!t]
\centering
\vspace{-0.2cm}
\makeatletter \if@twocolumn
\includegraphics[width=\columnwidth]{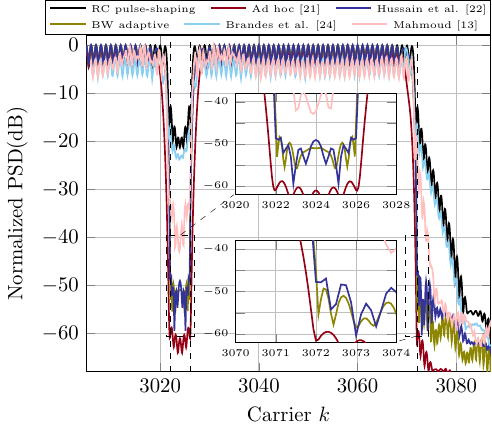}
\else
\includegraphics[width=8.5cm]{./Fig/TCOM_mask_Liter_methods_v4.pdf}
\fi \makeatother
\vspace{-0.7cm}
\caption{{\color{reviewer1}Normalized PSD attained by the proposed method and others taken from the literature. For those that use the proposed pulses, only $h_k(\mathrm{CC},t_k\text{-h})$ is considered.}}
\label{fig:MaskTCOMliterMethods}
\vspace{-0.8cm}
\end{figure}

Fig. \ref{fig:MaskTCOMliterMethods} shows the normalized \ac{PSD} attained in the aforesaid scenario. The \ac{RC} pulse-shaping case is shown as a reference. It can be observed that reducing the \ac{OOBE} in the spectral hole is more challenging than in the sideband. In the former, the \textit{bandwidth adaptive} method gives $\textrm{PSD}_{\textrm{max}}=-48$ dB , outperforming the technique in \cite{Mahmoud08} by $14.93$ dB and the one in \cite{Brandes10} by more than $20$ dB. {\color{reviewer1} It attains similar emissions to the method in \cite{Hussain2022}. As expected, the latter performs worse than the \textit{ad hoc} method, which generalizes \cite{Hussain2022} by optimizing the transition pulse to be used in each carrier (instead of optimizing a single one for all carriers).} The performance loss of the \textit{bandwidth adaptive} method with respect to the \textit{ad hoc} one is about $10.3$ dB. However, this loss is paid in exchange for an easy recomputation of the solutions whenever the spectral emission mask changes. Hence, if the notched band between carriers $3022$ and $3026$ is to be dynamically created without prior knowledge of the receiver, the proposed method determines the optimized pulses to be used by applying simple transformations to the ones already employed in the data carriers closest to the rightmost edge of the band (carriers $3056$ to $3068$), while costly optimization problems have to be solved online in the \textit{ad hoc} one. In the notched band starting in carrier $3072$, the \textit{bandwidth adaptive} method attains $\textrm{PSD}_{\textrm{max}}=-49$ dB, which is $24$ dB lower than the one given by \cite{Mahmoud08} {\color{reviewer1} and roughly $2.2$ dB lower than the one in \cite{Hussain2022}. Note that the \textit{bandwidth adaptive} slightly outperforms \cite{Hussain2022}, even though the pulses of the latter are optimized for the considered spectral mask and in the former they are obtained by adapting a precomputed solution. The obtained performance improvement is due to the additional degree of freedom, with respect to \cite{Hussain2022}, given by the use of a different transition pulse in each of the $N_{\textrm{h}}$ carriers.}

It must be mentioned that, in the considered scenario, the \textit{bandwidth adaptive} can achieve $\textrm{PSD}_{\textrm{max}}$ below $-50$ dB by increasing $N_{\textrm{h\_min}}=10$, at the cost of penalizing its performance in passband widths narrower than $14$ carriers. Finally, it must be highlighted that the \ac{PAPR} of the signal designed with the \textit{bandwidth adaptive} method is similar to that obtained with the \textit{ad hoc} solution, being the latter lower than the one resulting with \cite{Mahmoud08} and \cite{Brandes10} \cite[Fig. 6(b)]{Diez19}.

\section{Conclusion}
\label{Conclusion}
This work has proposed a framework to perform the spectral shaping of \ac{OFDM} signals subject to an emission mask that changes dynamically. The proposed method combines \ac{CC} and \ac{AST}. It firstly obtains the set of pulses to be applied to the data carriers of signals with very narrow passbands, which is the worst case from the \ac{OOBE} perspective. Low-complexity transformations are then proposed to dynamically adapt the precomputed pulses to changes in  the emission mask.  

The resulting method is computationally simple, transparent for the receiver, does not increase the \ac{PAPR} and allows the analytical calculation of the resulting \ac{PSD}. Numerical results show it achieves maximum normalized \ac{PSD} values below $-44$ dB in transmission bandwidths consisting of only $9$ carriers and below $-50$ dB in wider ones. 


%

\appendices

\section{}
\label{Apendice}
The proof of the third equality in (\ref{eq:localOpt:identidadesDesplazamiento}) is now given. Since the notched band $\mathcal{B}_n^+(l)$ starts at carrier $l$ and has normalized frequency span $B_{\textrm{n}}$, the $L \times L$ matrix $\boldsymbol{\Phi}_{\mathcal{B}_n^+(l)}$ can be expressed as
\begin{equation}
\makeatletter
\if@twocolumn
    \boldsymbol{\Phi}_{\mathcal{B}_n^+(l)} = \int_{\mathcal{B}_n^+(l)} \!\!\! \mathbf{f}_L(f) \mathbf{f}_L^{\textrm{H}}(f) df = \int_{\frac{l}{N}-B_{\textrm{n}}}^{\frac{l}{N}} \!\!\! \mathbf{f}_L(f) \mathbf{f}_L^{\textrm{H}}(f) df.
\else
    \boldsymbol{\Phi}_{\mathcal{B}_n^+(l)} = \int_{\mathcal{B}_n^+(l)} \mathbf{f}_L(f) \mathbf{f}_L^{\textrm{H}}(f) df = \int_{\frac{l}{N}-B_{\textrm{n}}}^{\frac{l}{N}} \mathbf{f}_L(f) \mathbf{f}_L^{\textrm{H}}(f) df.
\fi 
\makeatother
    \label{eq:appendix:definicionPhi}
\end{equation}

In the same fashion, 
\begin{equation}
\makeatletter
\if@twocolumn
    \begin{aligned}
        &\boldsymbol{\Phi}_{\mathcal{B}_n^+(l+\Delta k)} = \int_{\mathcal{B}_n^+(l+\Delta k)} \mathbf{f}_L(f) \mathbf{f}_L^{\textrm{H}}(f) df =\\ &=\int_{\frac{l+\Delta k}{N}-B_{\textrm{n}}}^{\frac{l+\Delta k}{N}} \mathbf{f}_L(f) \mathbf{f}_L^{\textrm{H}}(f) df,
    \end{aligned}
\else
    \boldsymbol{\Phi}_{\mathcal{B}_n^+(l+\Delta k)} = \int_{\mathcal{B}_n^+(l+\Delta k)} \mathbf{f}_L(f) \mathbf{f}_L^{\textrm{H}}(f) df = \int_{\frac{l+\Delta k}{N}-B_{\textrm{n}}}^{\frac{l+\Delta k}{N}} \mathbf{f}_L(f) \mathbf{f}_L^{\textrm{H}}(f) df,
\fi
\makeatother
    \label{eq:appendix:definicionPhiDesplaz}
\end{equation}
which by performing the change of the integration variable $f'=f-\frac{\Delta k}{N}$, yields
\begin{equation}
\makeatletter
\if@twocolumn
    \boldsymbol{\Phi}_{\mathcal{B}_n^+(l+\Delta k)} = \int_{\frac{l}{N}-B_{\textrm{n}}}^{\frac{l}{N}} \mathbf{f}_L(f'+\frac{\Delta k}{N}) \mathbf{f}_L^{\textrm{H}}(f'+\frac{\Delta k}{N}) df'.
\else
    \boldsymbol{\Phi}_{\mathcal{B}_n^+(l+\Delta k)} = \int_{\frac{l}{N}-B_{\textrm{n}}}^{\frac{l}{N}} \mathbf{f}_L(f'+\frac{\Delta k}{N}) \mathbf{f}_L^{\textrm{H}}(f'+\frac{\Delta k}{N}) df'.
\fi
\makeatother
    \label{eq:appendix:definicionPhiDesplaz2}
\end{equation}
Since $\mathbf{f}_L^{\textrm{H}}(f) = \left[ 1, e^{-j2\pi f}, \dots, e^{-j2\pi f(L-1)} \right]$,  
\makeatletter \if@twocolumn
\begin{equation}
    \begin{aligned}
        &\mathbf{f}_L^{\textrm{H}}(f'+\frac{\Delta k}{N}) = \left[ 1, e^{-j2\pi \left( f'+\frac{\Delta k}{N} \right)}, \dots, e^{-j2\pi \left( f'+\frac{\Delta k}{N} \right)(L-1)} \right] =\\ &= \mathbf{f}_L^{\textrm{H}}(f') \textrm{diag} \left\{ e^{-j\frac{2\pi}{N} \Delta k(0)}, \dots, e^{-j\frac{2\pi}{N} \Delta k  (L-1)} \right\}.
    \end{aligned}
    \label{eq:appendix:definicionFrecuenciaDeplaz}
\end{equation}
\else
\begin{equation}
    \begin{aligned}
        &\mathbf{f}_L^{\textrm{H}}(f'+\frac{\Delta k}{N}) = \left[ 1, e^{-j2\pi \left( f'+\frac{\Delta k}{N} \right)}, \dots, e^{-j2\pi \left( f'+\frac{\Delta k}{N} \right)(L-1)} \right] = \\ &= \mathbf{f}_L^{\textrm{H}}(f') \textrm{diag} \left\{ 1, e^{-j\frac{2\pi}{N} \Delta k }, \dots, e^{-j\frac{2\pi}{N} \Delta k  (L-1)} \right\}.
    \end{aligned}
    \label{eq:appendix:definicionFrecuenciaDeplaz}
\end{equation}
\fi \makeatother
    
Defining $\boldsymbol{\Omega}_{\Delta k}^{\textrm{H}}$ as 
\makeatletter \if@twocolumn
\begin{equation}
    \begin{aligned}
        &\boldsymbol{\Omega}_{\Delta k}^{\textrm{H}} = \textrm{diag} \left\{ e^{-j\frac{2\pi}{N} \Delta k (0-N_{\textrm{GI}})}, \dots, e^{-j\frac{2\pi}{N} \Delta k  (L-1-N_{\textrm{GI}})} \right\} =\\ &= \textrm{diag} \left\{ e^{-j\frac{2\pi}{N} \Delta k (0)}, \dots, e^{-j\frac{2\pi}{N} \Delta k  (L-1)} \right\} e^{j\frac{2\pi}{N}\Delta k N_{\textrm{GI}}}
    \end{aligned},
        \label{eq:appendix:descomposicionOmega}
\end{equation}
\else
\begin{equation}    
  \begin{aligned}
        &\boldsymbol{\Omega}_{\Delta k}^{\textrm{H}} = \textrm{diag} \left\{ e^{-j\frac{2\pi}{N} \Delta k (0-N_{\textrm{GI}})}, \dots, e^{-j\frac{2\pi}{N} \Delta k  (L-1-N_{\textrm{GI}})} \right\} =\\ &= \textrm{diag} \left\{ 1, e^{-j\frac{2\pi}{N} \Delta k }, \dots, e^{-j\frac{2\pi}{N} \Delta k  (L-1)} \right\} e^{j\frac{2\pi}{N}\Delta k N_{\textrm{GI}}},
    \end{aligned}
        \label{eq:appendix:descomposicionOmega}
\end{equation}
\fi \makeatother
yields $\mathbf{f}_L^{\textrm{H}}(f'+\frac{\Delta k}{N}) = \mathbf{f}_L^{\textrm{H}}(f') \boldsymbol{\Omega}_{\Delta k}^{\textrm{H}} e^{-j\frac{2\pi}{N}\Delta k N_{\textrm{GI}}}$. Substituting this in (\ref{eq:appendix:definicionPhiDesplaz2}) and simplifying leads to
\begin{equation}
\makeatletter
\if@twocolumn
    \begin{aligned}
        \boldsymbol{\Phi}_{\mathcal{B}_n^+(l+\Delta k)} = \boldsymbol{\Omega}_{\Delta k} \int_{\frac{l}{N}-B_{\textrm{n}}}^{\frac{l}{N}}   \mathbf{f}_L(f') \mathbf{f}_L^{\textrm{H}}(f') df' \boldsymbol{\Omega}_{\Delta k}^{\textrm{H}} = \boldsymbol{\Omega}_{\Delta k} \boldsymbol{\Phi}_{\mathcal{B}_n^+(l)} \boldsymbol{\Omega}_{\Delta k}^{\textrm{H}} ,
\end{aligned}
\else
        \boldsymbol{\Phi}_{\mathcal{B}_n^+(l+\Delta k)} = \boldsymbol{\Omega}_{\Delta k} \int_{\frac{l}{N}-B_{\textrm{n}}}^{\frac{l}{N}}   \mathbf{f}_L(f') \mathbf{f}_L^{\textrm{H}}(f') df' \boldsymbol{\Omega}_{\Delta k}^{\textrm{H}} = \boldsymbol{\Omega}_{\Delta k} \boldsymbol{\Phi}_{\mathcal{B}_n^+(l)} \boldsymbol{\Omega}_{\Delta k}^{\textrm{H}} ,
\fi
\makeatother
    \label{eq:appendix:definicionPhiDesplaz3}
\end{equation}
which equals the third equality in (\ref{eq:localOpt:identidadesDesplazamiento}).

\section*{Acknowledgment}
The authors would like to thank Maxlinear Hispania S.L. for the motivation of this work.
\ifCLASSOPTIONcaptionsoff
  \newpage
\fi



%
%
%
%

\bibliography{biblio}
\bibliographystyle{IEEEtran}

%




\end{document}